\documentclass[journal]{IEEEtran}
\usepackage{amsmath,amsfonts}
\usepackage{algorithmic}
\usepackage{algorithm}
\usepackage{array}
\usepackage[caption=false,font=normalsize,labelfont=sf,textfont=sf]{subfig}
\usepackage{textcomp}
\usepackage{stfloats}
\usepackage{url}
\usepackage{verbatim}
\usepackage{graphicx}
\usepackage{cite}
\usepackage{mathtools}
\usepackage{multirow}
\usepackage{paralist}
\usepackage{enumitem}
\usepackage[hyperfootnotes=false,bookmarks=false]{hyperref}
\usepackage{soul}
\usepackage{wasysym}
\usepackage{xspace}
\usepackage{siunitx}
\usepackage{nicefrac}
\usepackage{booktabs}
\usepackage{acronym}
\usepackage{units}

\acrodef{mir}[MIR]{Music Information Retrieval}
\acrodef{pcp}[PCP]{Pitch Class Profile}
\acrodef{dnn}[DNN]{Deep Neural Network}
\acrodef{cnn}[CNN]{Convolutional Neural Network}
\acrodef{rnn}[RNN]{Recurrent Neural Network}
\acrodef{ctc}[CTC]{Connectionist Temporal Classification}
\acrodef{mctc}[MCTC]{Multi-label CTC}
\acrodef{mfcc}[MFCC]{Mel Frequency Cesptral Coefficient}
\acrodef{stft}[STFT]{Short Time Fourier Transform}
\acrodef{cqt}[CQT]{Constant-Q-Transform}
\acrodef{cnn}[CNN]{Convolutional Neural Network}
\acrodef{gmm}[GMM]{Gaussian Mixture Model}
\acrodef{mlp}[MLP]{Multilayer Perceptron}
\acrodef{bce}[BCE]{Binary Cross-Entropy}
\acrodef{ce}[CE]{Categorical Cross-Entropy}
\acrodef{mse}[MSE]{Mean Square Error}
\acrodef{ace}[ACE]{Automatic Chord Recognition}
\acrodef{hmm}[HMM]{Hidden Markov Model}
\acrodef{mpe}[MPE]{Multi-Pitch Estimation}

\acrodef{amt}[AMT]{Automatic Music Transcription}
\acrodef{mpe}[MPE]{Multi-Pitch Estimation}
\acrodef{mfe}[MFE]{Multi-F0 estimation}
\acrodef{hcqt}[HCQT]{Harmonic Constant-Q Transform}
\acrodef{nn}[NN]{Neural Network}
\acrodef{asr}[ASR]{Automatic Speech Recognition}
\acrodef{cossim}[CS]{Cosine Similarity}
\acrodef{avgprec}[AP]{Average Precision}
\acrodef{acc}[Acc]{Accuracy}
\acrodef{erm}[ERM]{Empirical Risk Minimization}

\def\ap{\ac{avgprec}}

\def\etal{\emph{et al.}\xspace}

\def\schubert{\texttt{SWD}\xspace}
\def\musicnet{\texttt{MuN}\xspace}

\def\bach10{\texttt{B10}\xspace}
\def\trios{\texttt{Tri}\xspace}
\def\phenicx{\texttt{PhA}\xspace}
\def\choralsd{\texttt{CSD}\xspace}

\def\testThree{\texttt{MuN-3}\xspace}
\def\testTen{\texttt{MuN-10}\xspace}
\def\testTenWrong{\texttt{MuN-10}$_\mathrm{a}$\xspace}
\def\testTenSlow{\texttt{MuN-10}$_\mathrm{b}$\xspace}
\def\testTenFast{\texttt{MuN-10}$_\mathrm{c}$\xspace}
\def\testTenFull{\texttt{MuN-10}$_\mathrm{full}$\xspace}

\def\cnnNoArg{CNN\xspace}
\def\dcnnNoArg{DCNN\xspace}
\def\drcnnNoArg{DRCNN\xspace}
\def\unetNoArg{Unet\xspace}
\def\saunetNoArg{SAUnet\xspace}
\def\sausnetNoArg{SAUSnet\xspace}
\def\blunetNoArg{BLUnet\xspace}
\def\punetNoArg{PUnet\xspace}
\newcommand{\cnn}[1]{\cnnNoArg:#1}
\newcommand{\dcnn}[1]{\dcnnNoArg:#1}
\newcommand{\drcnn}[1]{\drcnnNoArg:#1}
\newcommand{\unet}[1]{\unetNoArg:#1}
\newcommand{\saunet}[1]{\saunetNoArg:#1}
\newcommand{\sausnet}[1]{\sausnetNoArg:#1}
\newcommand{\blunet}[1]{\blunetNoArg:#1}
\newcommand{\punet}[1]{\punetNoArg:#1}

\newcommand{\secref}[1]{\mbox{Section~\ref{#1}}}
\newcommand{\tabref}[1]{\mbox{Table~\ref{#1}}}
\newcommand{\figref}[1]{\mbox{Fig.~\ref{#1}}}

\begin{document}



\title{Deep-Learning Architectures for Multi-Pitch Estimation: Towards Reliable Evaluation}

\author{Christof Wei{\ss}, Geoffroy Peeters

\thanks{This work has been submitted to the IEEE for possible publication. Copyright may be transferred without notice, after which this version may no longer be accessible. The authors are with LTCI, T{\'e}l{\'e}com Paris, Institut Polytechnique de Paris, France. 
}
}

\markboth{ }
{Wei{\ss} \& Peeters: Deep-Learning Architectures for Multi-Pitch Estimation: Towards Reliable Evaluation}


\maketitle

\begin{abstract}
Extracting pitch information from music recordings is a challenging but important problem in music signal processing. 
Frame-wise transcription or multi-pitch estimation aims for detecting the simultaneous activity of pitches in polyphonic music recordings and has recently seen major improvements thanks to deep-learning techniques, with a variety of proposed network architectures. In this paper, we realize different architectures based on convolutional neural networks, the U-net structure, and self-attention components. We propose several modifications to these architectures including self-attention modules for skip connections, recurrent layers to replace the self-attention, and a multi-task strategy with simultaneous prediction of the degree of polyphony. We compare variants of these architectures in different sizes for multi-pitch estimation, focusing on Western classical music beyond the piano-solo scenario using the MusicNet and Schubert Winterreise datasets. Our experiments indicate that most architectures yield competitive results and that larger model variants seem to be beneficial. However, we find that these results substantially depend on randomization effects and the particular choice of the training--test split, which questions the claim of superiority for particular architectures given only small improvements. We therefore investigate the influence of dataset splits in the presence of several movements of a work cycle (cross-version evaluation) and propose a best-practice splitting strategy for MusicNet, which weakens the influence of individual test tracks and suppresses overfitting to specific works and recording conditions. A final evaluation on a mixed dataset suggests that improvements on one specific dataset do not necessarily generalize to other scenarios, thus emphasizing the need for further high-quality multi-pitch datasets in order to reliably measure progress in music transcription tasks.
\end{abstract}

\begin{IEEEkeywords}
Music information retrieval, music transcription, U-net, generalization, cross-version evaluation.
\end{IEEEkeywords}

\section{Introduction}\label{sec:introduction}
%
%
%
%
\ac{amt} is a central problem in signal processing and \ac{mir}, dealing with the conversion of music audio recordings into an explicit, human-readable representation of pitch, rhythm, and other musical information \cite{BenetosDDE19_TranscriptionOverview_SPM}. \ac{amt} is considered an ``enabling technology'' since it often serves as front-end for music analysis or music learning applications \cite{BenetosDDE19_TranscriptionOverview_SPM}. Because of the tremendous complexity of polyphonic pieces (such as a Romantic symphony), \ac{amt} remains challenging both as an algorithmic and a cognitive task---it requires extensive training for humans to acquire transcription skills. 
While a comprehensive \ac{amt} system for Western music aims for generating a human-readable score \cite{RomamPC18_AudioScoreTranscription_ISMIR}, several intermediate steps are considered 
and often developed independently, dealing with \emph{frame-level}, \emph{note-level}, \emph{stream-level}, or \emph{notation-level} transcription \cite{BenetosDDE19_TranscriptionOverview_SPM}. 
In this paper, we focus on \emph{frame-level transcription} or \ac{mpe}, i.\,e., detecting the simultaneous activity of a variable number of pitches over the course of an audio recording, which can be visualized in a piano-roll representation. \ac{mpe} is an important sub-task that needs to be under control for further advancements of \ac{amt}. 
In this paper, we consider the \emph{polyphonic} (multiple simultaneous pitches), \emph{polytimbral} (multiple simultaneous instruments), and \emph{instrument-agnostic} scenario, where we do not transcribe the source instrument (as opposed to, e.\,g., \cite{WuCS20_TranscriptionSelfattn_TASLP}). In particular, we want to overcome the frequent limitation to piano-solo music \cite{VincentBB10_SpectralDecomposition_TASLP,KelzDKBAW16_FramewisePianoTrans_ISMIR,HawthorneESRSRE18_OnsetsFrames_ISMIR,HawthorneSRSHDE19_MAESTRO_ICLR,HawthorneSSME21_PianoTranscrTransformers_ISMIR} and aim for a \emph{general-purpose} \ac{mpe} system, applicable for a wide range of Western classical music scenarios.

\begin{figure}[t!]
    \centering
    \includegraphics[width=.96\columnwidth]{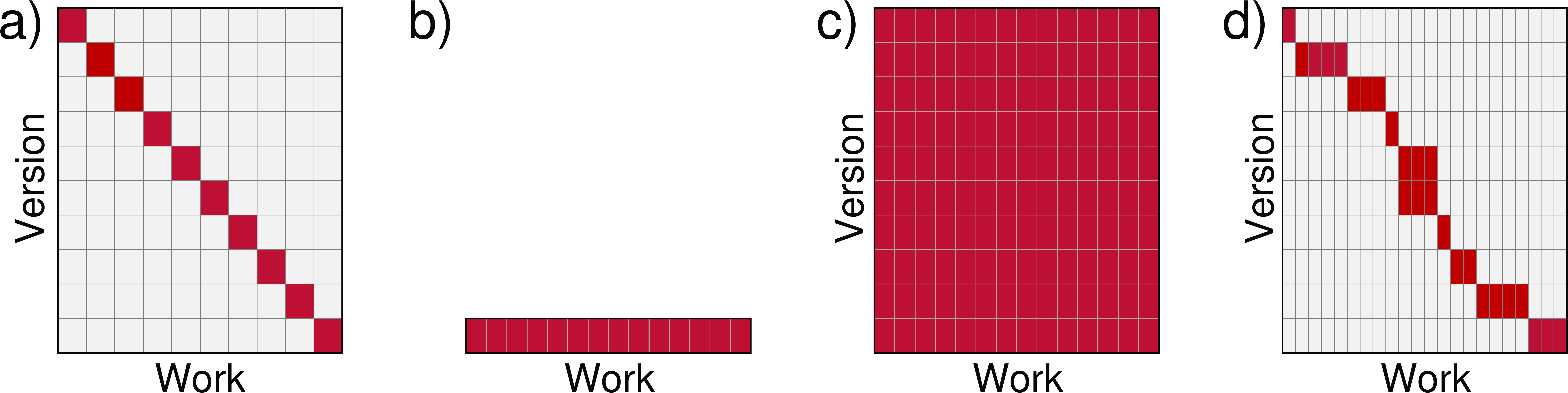}
    \vspace{-0.2cm}
    \caption{Types of music datasets. 
    \textbf{(a)~Separate-version dataset:} each work recorded in a different version. 
    \textbf{(b)~Single-version dataset:} all works recorded in the same version. 
    \textbf{(c)~Structured multi-version dataset:} each work recorded in the same set of versions. 
    \textbf{(d)~Mixed-version dataset:} some works recorded in the same version and/or some works recorded in several versions.
    }
    \label{fig:teaser}
\end{figure}

\textbf{Aspects of generalization.} 
Proving and improving such general applicability poses a challenge for any empirical experiment, requiring the variety of development data to be drawn from the same distribution of examples as the variety of intended application cases. 
To frame our \ac{mpe} problem in terms of Empirical Risk Minimization (ERM) \cite{Vapnik91_EmpiricalRiskMinimization_NIPS}, we assume the existence of an unknown joint probability $P(x,y)$ between the input $x\in X$ (the audio signal) and $y \in Y$ (the pitches).
Our goal is to estimate a function $\hat{y}\!=\!f_{\theta}(x)$ that approximates $P(y|x)$.
The difference between $y$ and $\hat{y}$ is measured using a loss $\mathcal{L}(f_{\theta}(x), y)$.
Here, $f$ defines a family of functions (in our case, one of the specific network architectures described in \secref{sec:ownmethods}) and $\theta$ its parameters (estimated with a variant of the SGD algorithm with the aim to minimize $\mathcal{L}$).
Since $P(x,y)$ is unknown, the risk $\mathcal{L}$ is empirically minimized over instances $(x_i,y_i)$ drawn from $P(x,y)$.
These instances should be chosen with care to guarantee the generalization of the learned $f_{\theta}$ to the whole sets $X$ and $Y$.
This generalization is often measured by separating a training, validation, and test subset---each drawn from the same $P(x,y)$. In practice, this separation poses the risk of unrealistic similarities between these subsets (e.\,g. the same works being present in training and test set), which must be avoided in order to prevent the models from unwanted overfitting. 
We therefore deal with three sources of variability:
(1) the choice of the family of functions $f$ (architectures),
(2) the estimation of the parameters $\theta$ (which depends not only on the optimization algorithm used but also on randomization aspects such as the initialization of $\theta$), and
(3) the choice of the instances $(x_i,y_i)$ drawn from $P(x,y)$ (training--test splits).
Most published works on MPE only study the influence of (1), claiming $f$ to be ``state-of-the-art'' without considering the influence of (2) and (3). In this paper, we reflect on such generalization issues for \ac{mpe}, thus emphasizing the importance of (2) and (3), which can drastically change conclusions related to (1).
%
%
%

\textbf{Scenario.} 
To this end, we focus on Western classical music scenarios, which we consider interesting for two main reasons: First, musical scores and recorded performances highly correspond to each other regarding pitch information, which allows for generating annotations in a semi-automatic way (see \secref{sec:datasets}). Second, musical compositions are usually accessible in various performances, which constitute separate interpretations and acoustic renderings, influenced by the performer style, recording conditions, audio mastering, and other aspects. We denote a specific setting of performers, instruments, and recording conditions as a \emph{version}: Usually, tracks on a CD (or compilation) are homogenous regarding these version aspects. In particular, this is the case for the individual \emph{works} (movements) of \emph{work cycle} (a multi-movement composition such as a sonata or string quartet). Previous research in \ac{mir} showed that not reflecting version or work similarities across training and test sets can lead to severe evaluation problems \cite{Sturm13_Accuracy_JIIS}, resulting in overfitting to either recording conditions (the so-called ``album effect'' in genre classification \cite{Flexer07_GenreClassification_ISMIR}) or to musical characteristics of works (a ``song effect'' for local key estimation \cite{WeissSM20_LocalKey_TASLP}).

\textbf{Multi-version datasets.} 
The presence or absence of several such versions characterizes different types of \ac{mpe} datasets (see \figref{fig:teaser}): Ideally, each work in a dataset is rendered in a different version (\figref{fig:teaser}a). In practice, however, many datasets are homogenous regarding version characteristics (single version, \figref{fig:teaser}b): Examples are Bach10 \cite{DuanPZ10_MultiF0_TASLP}, TRIOS \cite{FritschP13_ScoreInformedSourceSepNMFAndSynth_ICASSP}, PHENICX-Anechoic \cite{MironCBGJ16_SourceSepOrchestra_JECE}, or the Choral Singing Dataset \cite{CuestaGML18_ChoirIntonation_ICMPC}. In contrast, dedicated multi-version datasets (\figref{fig:teaser}c) such as the Schubert Winterreise Dataset \cite{WeissZAGKVM21_WinterreiseDataset_ACM-JOCCH} provide several versions of all works in a comprehensive fashion. Many datasets of classical music, however, are unstructured in this respect (\figref{fig:teaser}d), partially providing multiple works in the same version 
and/or multiple versions of the same work. This is the case for the MusicNet dataset \cite{ThickstunHK17_MusicNet_ICLR}, a great resource of 330 license-free audio files with pre-synchronized pitch annotations. MusicNet crucially stimulated the advancement of \ac{mpe} beyond the piano scenario \cite{ThickstunHFK18_Transcription_ICASSP,WuCS19_PolyphTranscription_ICASSP,CheukHS21_SemiSupervisedAMT_ACMMM} and has become the benchmark dataset for beyond-piano \ac{amt}. For this reason, we consider it as our central evaluation scenario. As one main contribution, we go beyond the standard test set used for MusicNet and investigate model generalization by systematically comparing different training--test splits, drawing inspiration from cross-version experiments as performed in \cite{WeissSM20_LocalKey_TASLP}. To this end, we consider the Schubert Winterreise Dataset \cite{WeissZAGKVM21_WinterreiseDataset_ACM-JOCCH} as a second dataset that allows to systematically study generalization across versions, works, or both. We finally compile a larger and more diverse dataset from several source datasets with different instrumentations, thus providing more insights into the robustness of the models.

\textbf{Deep-learning methods.} 
Such aspects of robustness and generalization are particularly important for \ac{mpe} systems based on supervised deep learning, which has become the de-facto standard approach \cite{KelzDKBAW16_FramewisePianoTrans_ISMIR,ThickstunHFK18_Transcription_ICASSP,HawthorneESRSRE18_OnsetsFrames_ISMIR,WuCS20_TranscriptionSelfattn_TASLP}. 
U-net models led to major advancements in \ac{mpe} \cite{WuCS19_PolyphTranscription_ICASSP,AbesserM21_BassTransUNet_Electronics,PedersoliTY20_TranscriptionPreStackUNet_ICASSP} and, recently, the inclusion of self-attention components \cite{WuCS20_TranscriptionSelfattn_TASLP,CheukHS21_SemiSupervisedAMT_ACMMM} or multi-task strategies \cite{WuCS19_PolyphTranscription_ICASSP,AbesserM21_BassTransUNet_Electronics} into U-nets showed further success. In this paper, we take up such network architectures based on U-net and self-attention components \cite{WuCS20_TranscriptionSelfattn_TASLP} and propose several extensions. While many of the mentioned studies claim state-of-the-art \ac{mpe} results, improvements usually happen in small increments, especially for beyond-piano scenarios. We questions this practice by addressing the results' dependency on randomization effects (initialization of parameters) and the models' generalization capabilities (training--test splits, cross-version and cross-dataset experiments). 

\textbf{Main contributions:} 
In summary, this paper focuses on the frame-wise, instrument-agnostic, polyphonic music transcription task (\ac{mpe}) beyond the piano-solo scenario using several public datasets, most importantly MusicNet \cite{ThickstunHK17_MusicNet_ICLR} and the Schubert Winterreise Dataset \cite{WeissZAGKVM21_WinterreiseDataset_ACM-JOCCH}. Our main contributions are the following: 
(1) We realize an enhanced U-net architecture with self-attention at the bottleneck layer---loosely inspired by \cite{WuCS20_TranscriptionSelfattn_TASLP}---and propose several modifications to this architecture including self-attention for skip connections, replacement of self-attention with recurrent layers, and multi-task training with additional degree-of-polyphony estimation at the bottleneck layer. 
(2) We compare variants of these models in different sizes among each other and to various baseline architectures, obtaining state-of-the-art results for several model variants. 
(3) We study the stability of results across different training runs and find that the training substantially depends on randomization effects, which questions previous conclusions about model superiority. 
(4) Inspired by this finding, we systematically compare the results' dependency on the choice of the training--test split and investigate generalization across works, versions, and datasets.
(5) Finally, we propose a best-practice evaluation strategy for MusicNet and other datasets, which is shown to be crucial for measuring progress in the \ac{mpe} task with data-driven systems.


The remainder of the paper is organized as follow: \secref{sec:datasets} discusses relevant datasets. 
\secref{sec:methods} summarizes related work on \ac{mpe} methods and sketches the models used in this paper. 
\secref{sec:experiments} presents the experiments. 
\secref{sec:discussion} discusses our findings. 
\secref{sec:conclusion} concludes the paper.

\section{Scenario and Datasets}\label{sec:datasets}

 \begin{table*}[t!]
\centering
\caption{Audio datasets with multi-pitch annotations used in this paper. 
$^1$\,For work cycles, each movement is counted as a work.
$^2$\,Versions per work for multi-version datasets.
} 
\footnotesize
\setlength{\tabcolsep}{5pt}
\renewcommand{\arraystretch}{0.9}
\vspace{-.2cm}
\begin{tabular}{lllllllr}
\toprule   
 ID & Dataset Name & Style / Instrumentation & Pitch annotation strategy & Mix Tracks & Works$^1$ & Versions$^2$ & hh:mm \\ 
\midrule
 \musicnet & MusicNet \cite{ThickstunHK17_MusicNet_ICLR} & Chamber music (piano, strings, winds) & Aligned scores &  330 & 306 & 1 up to 3 & 34:08 \\  
 \schubert & Schubert Winterreise \cite{WeissZAGKVM21_WinterreiseDataset_ACM-JOCCH} & Chamber music (piano, solo voice) & Aligned scores &  216 & 24 & 9 & 10:50 \\ 
 \trios & TRIOS \cite{FritschP13_ScoreInformedSourceSepNMFAndSynth_ICASSP} & Chamber music (piano, strings, winds) &  Multi-track &  5 & 5 & 1 &  0:03  \\   
 \bach10 & Bach10 \cite{DuanPZ10_MultiF0_TASLP} & Chamber music (violin, winds) & Multi-track &  10 & 10 & 1 & 0:06 \\ 
 \phenicx & PHENICX-Anechoic \cite{MironCBGJ16_SourceSepOrchestra_JECE} & Symphonic (orchestra)  & Multi-track &  4 & 4 & 1 & 0:10 \\ 
 \choralsd & Choral Singing Dataset \cite{CuestaGML18_ChoirIntonation_ICMPC} & A cappella (choir) &  MIDI-guided performance & 3 & 3 & 1 & 0:07  \\   
 \bottomrule
\end{tabular}
\label{tab:datasets}
\end{table*}
%
This section specifies the task approached by this paper and discusses the datasets used in this work (see \tabref{tab:datasets}).

\subsection{Task Definition}\label{sec:task} 
\begin{figure}[t!]
    \centering
    \includegraphics[width=.97\columnwidth]{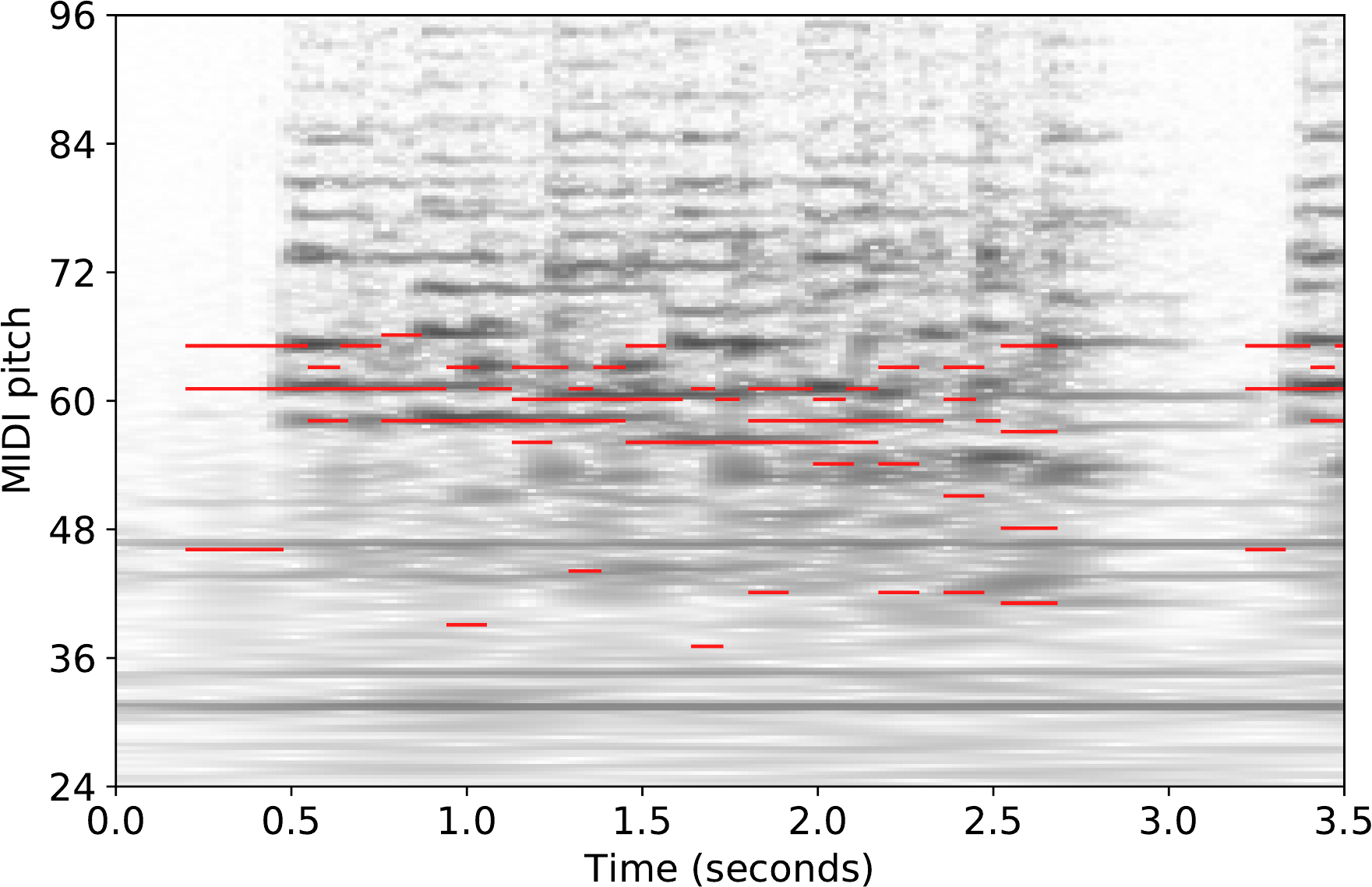}
    \vspace{-0.3cm}
     \caption{Id. 2382 from the MusicNet dataset, a recording of Beethoven's string quartet op.\,130, second movement, Presto. Log-compressed constant-Q transform (gray) and multi-pitch annotations (red) for the beginning.}
    \label{fig:mnetexample}
\end{figure}

This paper deals with the task of \ac{mpe} or \emph{frame-wise polyphonic music transcription} \cite{BenetosDDE19_TranscriptionOverview_SPM}, i.\,e., we aim for detecting the frame-wise activity of all pitches as indicated by a musical score, yet over the time axis of a recorded performance (audio). 
While this task is only a first step towards notation-level transcription \cite{BenetosDDE19_TranscriptionOverview_SPM}, the resulting piano-roll representation (red lines in \figref{fig:mnetexample}) is well-interpretable and constitutes a direct link to symbolic representations such as MIDI. As opposed to higher-level aspects of music notation (pitch spelling, instrument assignment), the \ac{mpe} subtask constitutes a more \emph{signal-level problem} and is closely related to multiple-fundamental-frequency or \ac{mfe}, which aims for extracting the detailed frequency trajectories of all harmonic sources in an audio signal, usually involving a sub-semitone frequency resolution (around \unit[10]{cents}) and a fine-grained time resolution (around \unit[10]{ms}). Beyond the resolution, pitch differs from F0 through being a perceptual phenomenon. It is not straightforward to map F0 to pitch in the presence of acoustic phenomena such as inharmonicities or vibrato (amplitudes up to several semitones in classical singing \cite{DriedgerBEM16_Vibrato_ISMIR}).

\subsection{Multi-Pitch Datasets} 
Successful training of high-capacity neural networks requires large amounts of annotated data. The limited availability of large datasets constitutes a major issue for \ac{mpe} \cite{BenetosDDE19_TranscriptionOverview_SPM}. Since a fully manual annotation is tedious and requires a high degree of expert knowledge, several strategies were proposed, as summarized by Su \etal \cite{SuY15_TranscriptionDatasets_CMMR}.

\textbf{Annotation strategies.} 
One strategy involves the use of MIDI-fied pianos for simultaneously generating audio and annotations, leading to piano datasets such as SMD \cite{MuellerKBA11_SMD_ISMIR-lateBreaking}, MAPS \cite{EmiyaBD10_MultipitchEstimation_TASLP}, or MAESTRO \cite{HawthorneSRSHDE19_MAESTRO_ICLR}. While this is limited to few instruments (piano, organ, synthesizers), Su \etal proposed an expert-in-the-loop approach to apply this strategy to non-piano music by re-playing parts on a MIDI instrument \cite{SuY15_TranscriptionDatasets_CMMR}.

Another strategy to go beyond the piano-solo scenario involves multi-track recordings to generate annotations with a monophonic F0-tracker on individual tracks. This was done, e.\,g., for MedleyDB \cite{BittnerSTMCB14_MedlyDB_ISMIR} and leads to multi-F0 rather than multi-pitch annotations (see \secref{sec:task}). Since the conversion of F0 to pitch is non-trivial, we do not use MedleyDB. 

However, multi-track recordings can also be used to simplify the manual pitch annotation process \cite{DuanPZ10_MultiF0_TASLP,FritschP13_ScoreInformedSourceSepNMFAndSynth_ICASSP,MironCBGJ16_SourceSepOrchestra_JECE}, which led to several small yet carefully-annotated \ac{mpe} datasets such as Bach10 (\bach10) \cite{DuanPZ10_MultiF0_TASLP}, TRIOS (\trios) \cite{FritschP13_ScoreInformedSourceSepNMFAndSynth_ICASSP}, or PHENICX-Anechoic (\phenicx) \cite{MironCBGJ16_SourceSepOrchestra_JECE} (all $\le$10 tracks, see \tabref{tab:datasets}). In other cases, the process is reversed: E.\,g., for the Choral Singing Dataset (\choralsd) \cite{CuestaGML18_ChoirIntonation_ICMPC}, the annotation (MIDI file) served as a guideline for the singers (together with a conductor video), which were recorded in groups according to the four musical parts. All these datasets have high-quality pitch annotations but were recorded specifically for research purposes in homogenous conditions and, therefore, constitute ``single-version'' datasets in the sense of \figref{fig:teaser}b.

\textbf{Annotation by score aligment.} 
As an alternative strategy, machine-readable scores can be exploited to generate pitch annotations, either using automated music synchronization \cite{MuellerZ21_SyncToolbox_JOSS} or relying on weakly-aligned training strategies using, e.\,g., the CTC loss \cite{WeissP21_MultiPitchMCTC_WASPAA}. 
For genres such as pop music, scores are rarely available or only roughly correspond to the pitch content of the recording, thus only serving as ``weak labels'' \cite{BenetosDDE19_TranscriptionOverview_SPM}. This is different for professional recordings of classical music, where a one-to-one correspondence between score and audio can be assumed (up to structural differences). 

%

Such score--audio pairs of classical music were considered to generate the MusicNet (\musicnet) dataset \cite{ThickstunHK17_MusicNet_ICLR}, an excellent resource of 330 license-free audio files with pre-synchronized pitch annotations that crucially stimulated the advancement of \ac{mpe} \cite{ThickstunHFK18_Transcription_ICASSP,WuCS19_PolyphTranscription_ICASSP,CheukHS21_SemiSupervisedAMT_ACMMM}.\footnote{Now hosted on Zenodo: \protect \url{https://zenodo.org/record/5120004}.} 
\musicnet is a very challenging dataset for \ac{mpe}, yielding consistently and substantially worse results (F-measure below 75\% \cite{WuCS19_PolyphTranscription_ICASSP}) across all studies compared to piano datasets such as MAPS or MAESTRO (up to 90\% F-measure \cite{HawthorneSRSHDE19_MAESTRO_ICLR}). We suspect three central reasons for this: First, as reported by several authors \cite{WuCS20_TranscriptionSelfattn_TASLP,GardnerSMHE21_MultiTaskTranscription_arXiv}, the annotations in \musicnet are generated with automated synchronization, which leads to inaccuracies or synchronization errors (shifts) that especially affect frames around the start and end of notes or works in fast tempo (see \figref{fig:mnetexample} for the latter). Second, \musicnet contains license-free audio files, which are often digitized from old records suffering from audio artifacts (in contrast to commercial recordings) and stem from a variety of sources, interpreters, and acoustic conditions (in contrast to the quite homogenous conditions of MAPS or MAESTRO). Third, \musicnet captures a great variety of instruments, genres, and composers, which are, however, quite imbalanced in the dataset (see \cite{ThickstunHK17_MusicNet_ICLR}): It mainly contains chamber music for solo instruments (piano, violin, cello), with piano solo accounting for almost half the total audio length, but also works for up to six players involving piano, strings, and wind instruments. Regarding composers, there is also an imbalance, with Beethoven accounting for more than half of the audio content. These imbalances hamper generalization to the underrepresented scenarios. In this paper, we investigate the latter issue by systematically comparing different training--test splits on \musicnet, thereby drawing inspiration from cross-version experiments as performed in \cite{WeissSM20_LocalKey_TASLP}. In this context, it is important to note that \musicnet is a mixed-version dataset as indicated by \figref{fig:teaser}d: Most often, several movements (usually two to five) of a work cycle are included (in a specific version), but it is unknown whether different cycles stem from the same version (are recorded under the same conditions). Some works also exist in multiple versions.\footnote{Multiple versions exist for movements from eleven different cycles. 
In total, 24 of the 330 tracks are other versions of a work already in the dataset. See our code repository for a detailed list.} However, the dataset is neither structured nor comprehensive in this respect.
Therefore, we further consider the Schubert Winterreise Dataset (\schubert) \cite{WeissZAGKVM21_WinterreiseDataset_ACM-JOCCH} as a comprehensive multi-version dataset (\figref{fig:teaser}c), which allows to systematically study generalization across versions, works, or both, while being restricted to a smaller and more homogenous set of musical works (the 24 songs of the \emph{Winterreise} cycle). \schubert comprises nine recorded performances (two free and seven commercial versions), scores, and pitch annotations,\footnote{\protect \url{https://zenodo.org/record/5139893}} generated with a synchronization algorithm \cite{MuellerZ21_SyncToolbox_JOSS}.

We finally compile a larger and more diverse dataset from several source datasets with different instrumentations (\tabref{tab:datasets}), allowing for insights into model behavior for a broader range of annotation strategies and music scenarios.

\section{Technical Methods}\label{sec:methods}
This section summarizes relevant deep-learning approaches to \ac{mpe} focusing on the beyond-piano case (\secref{sec:prevmethods}) and, in particular, on models tested on \musicnet. We then present the architectures considered in this paper (\secref{sec:ownmethods}).

\subsection{Previous Models}\label{sec:prevmethods}
Traditional approaches to \ac{mpe} typically rely on matrix factorization techniques \cite{VincentBB10_SpectralDecomposition_TASLP,BenetosD13_MusicTranscription_JASA}. Over the last years, deep-learning approaches led to major improvements for this task \cite{KelzDKBAW16_FramewisePianoTrans_ISMIR,ThickstunHFK18_Transcription_ICASSP,HawthorneESRSRE18_OnsetsFrames_ISMIR,WuCS20_TranscriptionSelfattn_TASLP}. In this context, a variety of different network architectures such as a \ac{rnn} \cite{SigtiaBD16_DNNPolyPianoTrans_TASLP}, a \ac{cnn} \cite{KelzDKBAW16_FramewisePianoTrans_ISMIR,ThickstunHFK18_Transcription_ICASSP}, or their combination \cite{HawthorneESRSRE18_OnsetsFrames_ISMIR,HawthorneSRSHDE19_MAESTRO_ICLR} were proposed. 
Kelz et al. \cite{KelzDKBAW16_FramewisePianoTrans_ISMIR} showed that a \ac{cnn} performs on par with dense or recurrent architectures while having fewer parameters---once it is used with a suitable input representation. A particular input representation is the \ac{hcqt}, where CQTs in harmonic frequency ratios are stacked depth-wise such that harmonic relationship is represented along the channel dimension \cite{BittnerMSLB17_DeepSalience_ISMIR}. Using \ac{hcqt} as input to a \ac{cnn} has shown success for \ac{mfe} \cite{BittnerMSLB17_DeepSalience_ISMIR} and \ac{mpe} \cite{WeissP21_MultiPitchMCTC_WASPAA}. For these reasons, we base our methods on convolutional blocks and use an \ac{hcqt} as input representation. As a major advancement, U-net models \cite{RonnebergerFB15_UNet_MICCAI} were shown to improve performance of \ac{amt} \cite{WuCS19_PolyphTranscription_ICASSP,WuCS20_TranscriptionSelfattn_TASLP,AbesserM21_BassTransUNet_Electronics,PedersoliTY20_TranscriptionPreStackUNet_ICASSP,CheukLBH20_SpecReconstructionAMT_ICPR,CheukHS21_SemiSupervisedAMT_ACMMM} and other \ac{mir} tasks \cite{StollerED18_WaveUNet_ISMIR,DorasEP19_MelodyUNet_MMRP}. More recently, the inclusion of self-attention components into U-nets \cite{WuCS20_TranscriptionSelfattn_TASLP,CheukHS21_SemiSupervisedAMT_ACMMM} and other models \cite{CheukLBH21_OnsetsFramesAttention_IJCNN} was applied successfully to \ac{mpe}. Most of our architectures rely on the U-net paradigm, enhanced with self-attention components as in \cite{WuCS20_TranscriptionSelfattn_TASLP} and other extensions.

\textbf{Multi-task strategies.} 
While the models mentioned above are usually trained on the \ac{mpe} task in isolation, major success was reported when proceeding to multi-task settings, most prominently the onset-and-frames approach \cite{HawthorneESRSRE18_OnsetsFrames_ISMIR,HawthorneSRSHDE19_MAESTRO_ICLR,CheukLBH21_OnsetsFramesAttention_IJCNN}. However, this strategy is tailored to the piano-solo case (where the percussive keystrokes help to successfully track onsets) and therefore, is not generally applicable to other instruments with less prominent onsets. More recently, major advancements were reported for other, more difficult transcription tasks such as note-level  \cite{HawthorneSRSHDE19_MAESTRO_ICLR,HawthorneSSME21_PianoTranscrTransformers_ISMIR,GardnerSMHE21_MultiTaskTranscription_arXiv,WuCS20_TranscriptionSelfattn_TASLP} and instrument-level transcription \cite{WuCS20_TranscriptionSelfattn_TASLP}, where self-attention components \cite{WuCS20_TranscriptionSelfattn_TASLP} and Transformer architectures \cite{VaswaniSPUJGKP17_Attention_NIPS} have been applied successfully.

%

\begin{figure*}[th!]
    \centering
    \includegraphics[width=.89\textwidth]{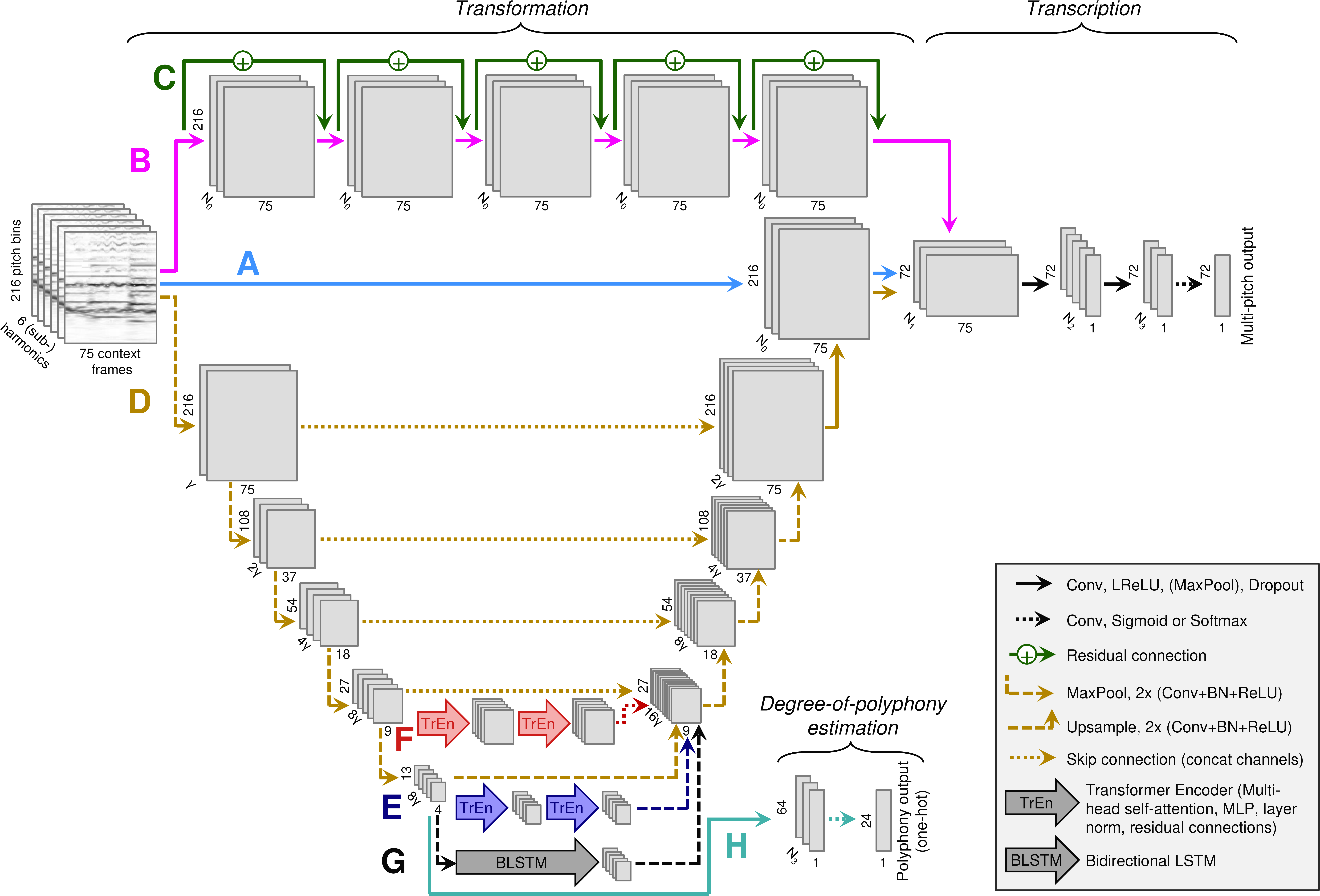}
     \vspace{-0.2cm}
    \caption{Overview of the model architectures used in this work. With the components (capital letters), we build several model types such as 
    CNNs (A), 
    Deep CNNs (B), 
    Deep Residual CNNs (B+C), 
    U-Nets (D), 
    U-Nets with self-attention at the bottleneck (D+E) and
    also at the lowest skip connection (D+E+F), 
    U-Nets with BLSTM (D+G), 
    and U-Nets with polyphony estimation (D+H). 
    See \tabref{tab:architecturesmnet} for details on parameters and sizes. [Best viewed in color.]
    }
    \label{fig:architectures}
\end{figure*}


\textbf{Methods evaluated on MusicNet.} 
Returning to the frame-level \ac{mpe} task, we now discuss relevant studies that focus on Western classical music beyond the piano-solo scenario and report results on the \musicnet dataset. By introducing \musicnet, Thickstun \etal propose several baseline methods working on raw audio input. Building on this, they further propose CNN-based architectures and data augmentation strategies (tuning and transposition) obtaining substantial improvements \cite{ThickstunHFK18_Transcription_ICASSP}.

Wu \etal \cite{WuCS19_PolyphTranscription_ICASSP} employed a U-Net with enhancement (atrous spatial pyramid pooling) for ``semantic segmentation''. As input representation, they consider a frequency--periodicity representation 
also including harmonics, similar to the \ac{hcqt} idea. They achieve an F-measure of 73.7\%, one of the best results on \musicnet so far. In \cite{WuCS20_TranscriptionSelfattn_TASLP}, they extend the task towards multi-instrument transcription, enhancing the model of \cite{WuCS19_PolyphTranscription_ICASSP} with self-attention at the bottleneck, with improved results for instrument-wise transcription but similar results on \ac{mpe}.

Another U-net-based \ac{mpe} approach was proposed by Pedersoli \etal \cite{PedersoliTY20_TranscriptionPreStackUNet_ICASSP}, who investigate the ``pre-stacking'' of a U-Net to different CNNs. Here, the U-Net serves as a ``transformation network'' followed by the CNN as a ``transcription network,'' which they found to be more effective than just increasing depth or width. In this paper, we take up their idea of pre-stacking the U-net. Cheuk \etal  \cite{CheukLBH20_SpecReconstructionAMT_ICPR} consider two U-Nets, one for transcription and one for spectrogram reconstruction, connected with Bidirectional LSTM components 
and a reconstruction loss. Aiming towards better model generalization, they combined the reconstruction paradigm with virtual adversarial training \cite{CheukHS21_SemiSupervisedAMT_ACMMM}. To put a semantic meaning to the bottom part (bottleneck), some authors  train the U-net in a multi-task fashion \cite{HsiehSY19_UNetMelody_ICASSP,AbesserM21_BassTransUNet_Electronics} by extracting a voice activity estimate from the bottleneck, applied for monophonic pitch estimation of the predominant melody \cite{HsiehSY19_UNetMelody_ICASSP} or the bass line \cite{AbesserM21_BassTransUNet_Electronics}.
 
Beyond the U-net, further architectures were tested on \musicnet. Steiner \etal  \cite{SteinerSBJ20_MultipitchEchoStateNetworks_EUSIPCO} test echo-state-networks, a kind of \ac{rnn}. Gardner \etal \cite{GardnerSMHE21_MultiTaskTranscription_arXiv} approach multi-instrument transcription tasks with Transformers. The obtained promising results, which dropped when using cross-dataset evaluation. Investigating \musicnet, they point to synchronization inaccuracies and labeling issues---aspects that we discuss later in this paper.

\subsection{Proposed Models}\label{sec:ownmethods}
For the experiments in this paper, we consider a variety of deep-learning approaches as summarized in \figref{fig:architectures}. Since describing all details in a comprehensive way is not possible here, we restrict ourselves to describing the central concepts and refer to our source code\footnote{Repository published upon acceptance, preview under \protect \url{https://gitfront.io/r/user-1366076/bd49d9799d8e4929f2d6afc436765d1251e76f2e/multipitch-architectures/}} for implementation details. All model variants share a convolutional post-processing stage as proposed by \cite{WeissP21_MultiPitchMCTC_WASPAA}, taking up the idea of a \emph{transformation} and a \emph{transcription} part \cite{PedersoliTY20_TranscriptionPreStackUNet_ICASSP}, which are trained together. As input, we use a log-compressed \ac{hcqt} with five harmonics and one subharmonic, comprising six octaves\footnote{While there are some individual notes in \musicnet going beyond that limit, the pitch distribution in \cite{ThickstunHK17_MusicNet_ICLR} shows that this is negligible in practice.} (C1--C7, or MIDI pitches 24--96) with three bins per semitone, resulting in 216 pitch bins, and using librosa's tuning estimation for adjusting bin frequencies. We rely on a sample rate of \unit[22.05]{kHz} and an \ac{hcqt} hopsize of 512 samples, resulting in a frame rate of \unit[43.07]{Hz}. All models operate on input patches of 75 frames (center frame plus 37 context frames to each side) corresponding to \unit[1.74]{s} of audio, processed by layer normalization, and predict the pitch activities of the center frame.


\textbf{\cnnNoArg architectures.} 
Our basic model variant is a \ac{cnn} with five layers (\figref{fig:architectures}\,A, blue path) as used in \cite{WeissP21_MultiPitchMCTC_WASPAA} for \ac{mpe} and in \cite{WeissZZSM21_DeepChromaChord_ISMIR} for pitch-class estimation. The first layer performs prefiltering (compression of the harmonics/channels) and preserves the input size, followed by reduction to pitches (one bin per semitone), temporal reduction, and a channel reduction. The model size (number of parameters) can be adjusted by increasing the width (number of channels/kernels) in each layer, specified by N$_0$, N$_1$, N$_2$, N$_3$ (see \tabref{tab:architecturesmnet} for concrete realizations). All layers use 2D-convolutions followed by LeakyReLU activations (slope 0.3), 2D-max-pooling, and dropout (rate 0.2). The final layer has a sigmoid activation since the models are trained with \ac{bce} loss. For the following model variants, we only replace the first (prefiltering) layer by more complex components, keeping all subsequent layers identical to \cnnNoArg.

\textbf{\dcnnNoArg and \drcnnNoArg architectures.} 
Inspired by \cite{WeissZZSM21_DeepChromaChord_ISMIR}, we realize a Deep \ac{cnn} (\dcnnNoArg) variant by replicating the pre-filtering layer five times (\figref{fig:architectures}\,B, magenta path). We enhance this variant with additive skip connections (residual connections) around the pre-filtering layers, resulting in a Deep Residual (\drcnnNoArg) architecture inspired by the ResNet \cite{HeZRS16_DeepResidualLerning_CVPR} (\figref{fig:architectures}\,B+C, magenta and green paths).

\textbf{\unetNoArg architectures}. 
Due to its frequent and successful use for \ac{mpe}, we consider the U-net \cite{RonnebergerFB15_UNet_MICCAI} as basis for our further architectures. We adopt the idea of pre-stacking a U-net \cite{PedersoliTY20_TranscriptionPreStackUNet_ICASSP} to our post-processing \ac{cnn} (\figref{fig:architectures}\,D, golden path). After an initial double-convolution layer with batch normalization (BN) and ReLU activation, we employ four downsampling steps (max-pooling by a factor of two) followed by the double-convolution module, and four upsampling steps followed by double-convolution. We realize the skip connections by concatenating channels from the downsampling and upsampling part. The width (channels) can be controlled with a parameter $\gamma$ leading to U-nets in different sizes (see \tabref{tab:architecturesmnet}d).

\textbf{\saunetNoArg and \sausnetNoArg (with self-attention components).} 
Inspired by \cite{WuCS20_TranscriptionSelfattn_TASLP}, we consider an extension of the U-net with self-attention at the bottom part (bottleneck) denoted as Self-Attention U-net (\saunetNoArg, \figref{fig:architectures}\,D+E, golden and navy paths). As opposed to the memory-flange self-attention in \cite{WuCS20_TranscriptionSelfattn_TASLP}, applied to patches of the downsampled representation via raster-scan, we simply flatten the time and frequency axes and use two Transformer encoder blocks as proposed by \cite{VaswaniSPUJGKP17_Attention_NIPS}. Such a block consists of a multi-head self-attention and a fully connected network (multi-layer perceptron). Both blocks are followed by dropout and layer normalization and applied in a residual fashion (output added to the previous input). As in \cite{VaswaniSPUJGKP17_Attention_NIPS}, we add absolute positional encodings as a combination of sine and cosine functions in order to inform the self-attention about the sequence order. In accordance with \cite{VaswaniSPUJGKP17_Attention_NIPS}, we obtained similar results when using learned positional encodings instead. We always use eight attention heads so that the embedding dimension is determined only by the U-net size $\gamma$. The number of hidden nodes is parametrized by $\lambda$. Among the different sizes listed in \tabref{tab:architecturesmnet}, we consider the \saunet{L} with roughly eight million parameters as most similar to the model presented in \cite{WuCS20_TranscriptionSelfattn_TASLP}. We further propose a modification to this architecture by introducing self-attention for the lowest skip connection (\sausnetNoArg, \figref{fig:architectures}\,D+E+F, golden, navy, and red paths), using the same Transformer encoders as described above. Beyond this paper's results, we also experimented with self-attention at higher skip connections obtaining worse results. 

\textbf{\blunetNoArg (with recurrent components).} 
Since the attention mechanism has been proposed for effectively processing sequence data \cite{VaswaniSPUJGKP17_Attention_NIPS}, we also test a recurrent variant where we replace self-attention by Bidirectional Long-Short-Term Memory (BLSTM) 
component (\blunetNoArg, \figref{fig:architectures}\,D+G, golden and black paths), the more common technique to capture sequence information. In this case, the parameter $\lambda$ denotes the number of hidden units in the BLSTM layer (see \tabref{tab:architecturesmnet}).

\textbf{\punetNoArg (with polyphony estimation)}. 
Finally, we consider a multi-task strategy to enforce a semantic meaning of the lowest U-net layer (bottleneck). We draw inspiration from \cite{HsiehSY19_UNetMelody_ICASSP,AbesserM21_BassTransUNet_Electronics}, where the bottleneck is trained with an auxiliary task: predicting the activity (voicing) in monophonic pitch estimation scenarios. Since inactive frames are rare in our multi-pitch scenario, we propose an alternative: As an auxiliary task, we predict the local \emph{degree of polyphony}, i.\,e., the number of active pitches in the center frame (0--23), which turned out to be a useful side information for \ac{mpe} \cite{DuanPZ10_MultiF0_TASLP,TaenzerMA21_PianoPolyphony_Electronics}. We derive this information from the bottleneck of the U-net, post-processed by a small, two-layer \ac{cnn} with 24-class softmax output. This variant is called the Polyphony U-net (\punetNoArg, \figref{fig:architectures}\,D+H, golden and aquamarine paths).

\section{Experiments and Results}\label{sec:experiments}
This section presents our comprehensive experiments and results. We first detail the experimental procedure (\secref{sec:procedure}). Next, we compare the performance of our architectures in varying size (number of parameters) on \musicnet (\secref{sec:modelsizes}). We report on the reliability of these results and investigate different test subsets of \musicnet (\secref{sec:testsets}). Then, we perform a systematic cross-version study on \schubert (\secref{sec:crossversion}). Finally, we test generalization to other scenarios by compiling a mixed dataset (\secref{sec:crossdataset}).
 
\subsection{Experimental Protocol}\label{sec:procedure}
We now describe the experimental protocol to train and test our models for \ac{mpe}. The details of the dataset splits are discussed later when presenting the respective results.
 
\textbf{Training procedure.} 
For training our models, we feed mini-batches of 25 HCQT excerpts as input, each having a size of 75 context frames, 216 pitch bins, and 6 (sub)-harmonics (see \secref{sec:ownmethods} for details) with slight logarithmic compression, along with the corresponding multi-pitch annotations for the center frame. We sample these excerpts from the full tracks' HCQT at regular intervals and adjust the stride parameter (hop size) such that the training set comprises roughly 95k training examples (3800 batches) and 9k validation examples (360 batches). All these examples are processed for one epoch to be finished. We train the model for a maximum of 100 epochs with the AdamW optimizer \cite{LoshchilovH19_AdamW_ICLR}, a stabilized version of the Adam optimizer using decoupled weight decay regularization, with an initial learning rate 0.001. For scheduling, we halve the learning rate if the validation loss plateaus for five epochs. After twelve non-improving epochs, we perform early stopping and use the best-performing model (with respect to the validation loss) for testing. In practice, we rarely need more than 20 epochs to reach convergence; only the simpler models (\cnnNoArg and \dcnnNoArg) require more epochs (up to 50). As criterion, we compute the \ac{bce} loss between pitch predictions and targets. For the \punetNoArg model, we also calculate the \ac{ce} loss for the degree-of-polyphony output (compare \figref{fig:architectures}) and sum the two losses. To have the losses in a similar order, we scale the \ac{ce} loss with a factor of 0.04 (empirically tested, not further optimized).

\begin{table}[t]
\caption{Comparison of network architectures, trained on the dataset \musicnet and tested on the test set \testTenWrong.} 
\vspace{-.2cm}
\centering
\footnotesize
\setlength{\tabcolsep}{3pt}
\renewcommand{\arraystretch}{0.96}
\begin{tabular}{lrrrr|llll}
\toprule
 Model & Param. & Channels & $\gamma$ & $\lambda$ & \multicolumn{4}{c}{Eval. Measures (\%)} \\
 Type:Size & $/10^3$ & N$_0$,N$_1$,N$_2$,N$_3$ &  &  & P & R & F & AP \\ 
\midrule
\multicolumn{9}{c}{\textbf{(a) \cnnNoArg} (\figref{fig:architectures}\,A)} \\
\midrule
\cnn{XS} & 48 & 20,20,10,1 & -- & -- & 69.0 & 61.4 & 64.5 & 69.3 \\ 
\cnn{S} & 603 & 100,100,50,10 & -- & -- & 68.7 & 67.8 & 68.0 & 73.2 \\ 
\cnn{M} & 1,813 & \!250,150,100,100 & -- & -- & 71.0 & 71.5 & 70.9 & 75.3 \\ 
\cnn{L} & 2,467 & \!280,180,120,100 & -- & -- & 70.4 & 71.7 & 70.7 & 75.0 \\ 
\midrule
\multicolumn{9}{c}{\textbf{(b) \dcnnNoArg} (\figref{fig:architectures}\,B)} \\
\midrule 
\dcnn{S} & 408 & 20,20,10,1 & -- & -- & 84.0 & 42.7 & 55.4 & 76.7 \\ 
\dcnn{M} & 1,602 & 40,40,30,10 & -- & -- & 68.2 & 80.9 & 73.8 & 78.6 \\ 
\dcnn{L} & 4,815 & 70,70,50,10 & -- & -- & 78.2 & 71.0 & 72.8 & 78.2  \\ 
\midrule
\multicolumn{9}{c}{\textbf{(c) \drcnnNoArg} (\figref{fig:architectures}\,B+C)} \\
\midrule
\drcnn{S} & 408 & 20,20,10,1 & -- & -- & 82.1 & 49.6 & 61.3 & 77.0 \\ 
\drcnn{M} & 1,602 & 40,40,30,10 & -- & -- & 70.3 & 78.2 & 73.9 & 78.5 \\ 
\drcnn{L} & 4,815 & 70,70,50,10 & -- & -- & 70.1 & 78.2 & 73.6 & 78.2 \\ 
\midrule
\multicolumn{9}{c}{\textbf{(d) \unetNoArg} (\figref{fig:architectures}\,D)} \\
\midrule
\unet{S} & 882 & 64,30,20,10 & 8 & -- & 72.1 & 75.3 & 73.3 & 77.8 \\ 
\unet{M} & 1,655 & 128,100,80,50 & 8 & -- & 71.9 & 76.5 & 73.9 & 78.0 \\ 
\unet{L} & 4,552 & 128,150,100,80 & 16 & -- & 73.9 & 74.1 & 73.8 & 79.0 \\ 
\unet{XL} & 14,252 & \!128,180,150,100 & 32 & -- & 71.9 & 77.9 & 74.6 & 79.1 \\ 
\midrule
\multicolumn{9}{c}{\textbf{(e) \saunetNoArg} (\figref{fig:architectures}\,D+E)} \\ 
\midrule
\saunet{M} & 1,180 & 64,30,20,10 & 8 & \!1024 & 72.4 & 73.3 & 72.6 & 77.0 \\ 
\saunet{L} & 7,983 & 128,80,50,30 & 16 & \!8192 & 73.7 & 76.6 & 75.0 & 79.8 \\ 
\saunet{XL} & 10,093 & \!128,200,150,150 & 16 & \!8192 & 71.5 & 77.6 & 74.2 & 78.8 \\ 
\saunet{XXL}\!\! & 23,439 & \!128,200,150,150 & 32 & \!8192 & 72.4 & 76.3 & 74.0 & 78.6 \\ 
\midrule
\multicolumn{9}{c}{\textbf{(f) \sausnetNoArg} (\figref{fig:architectures}\,D+E+F)} \\
\midrule
\sausnet{M} & 1,213 & 64,30,20,10 & 8 & 512 & 74.0 & 70.0 & 71.8 & 76.9 \\ 
\sausnet{L} & 8,115 & 128,80,50,30 & 16 & \!4096 & 75.8 & 71.6 & 73.4 & 79.0 \\ 
\sausnet{XL} & 14,436 & \!128,200,150,150 & 16 & \!8192 & 73.7 & 77.1 & 75.1 & 79.6 \\ 
\sausnet{XXL}\!\!\! & 32,371 & \!128,200,150,150 & 32 & \!8192 & 70.9 & 76.7 & 73.4 & 77.7 \\ 
\midrule
\multicolumn{9}{c}{\textbf{(g) \blunetNoArg} (\figref{fig:architectures}\,D+G)} \\
\midrule
\blunet{M} & 1,343 & 64,30,20,10 & 4 & 208 & 73.4 & 72.0 & 72.4 & 77.1 \\ 
\blunet{L} & 9,649 & 128,80,50,30 & 8 & 416 & 71.8 & 76.9 & 73.6 & 78.5 \\ 
\blunet{XXL}\! & 22,376 & \!128,200,150,150 & 16 & 832 & 73.2 & 73.6 & 73.1 & 77.9 \\ 
\midrule
\multicolumn{9}{c}{\textbf{(h) \punetNoArg} (\figref{fig:architectures}\,D+G)} \\
\midrule
\punet{M} & 1,680 & 128,100,80,50 & 8 & -- & 75.1 & 74.0 & 74.3 & 79.4 \\ 
\punet{L} & 4,643 & 128,100,80,50 & 16 & -- & 72.7 & 75.8 & 74.0 & 78.9 \\ 
\punet{XL} & 14,598 & \!128,180,150,100 & 32 & -- & 72.8 & 75.8 & 74.1 & 79.1 \\ 
\bottomrule
\end{tabular}
\label{tab:architecturesmnet}
\end{table}

\begin{figure}[t]
    \flushright
    \includegraphics[width=.97\columnwidth]{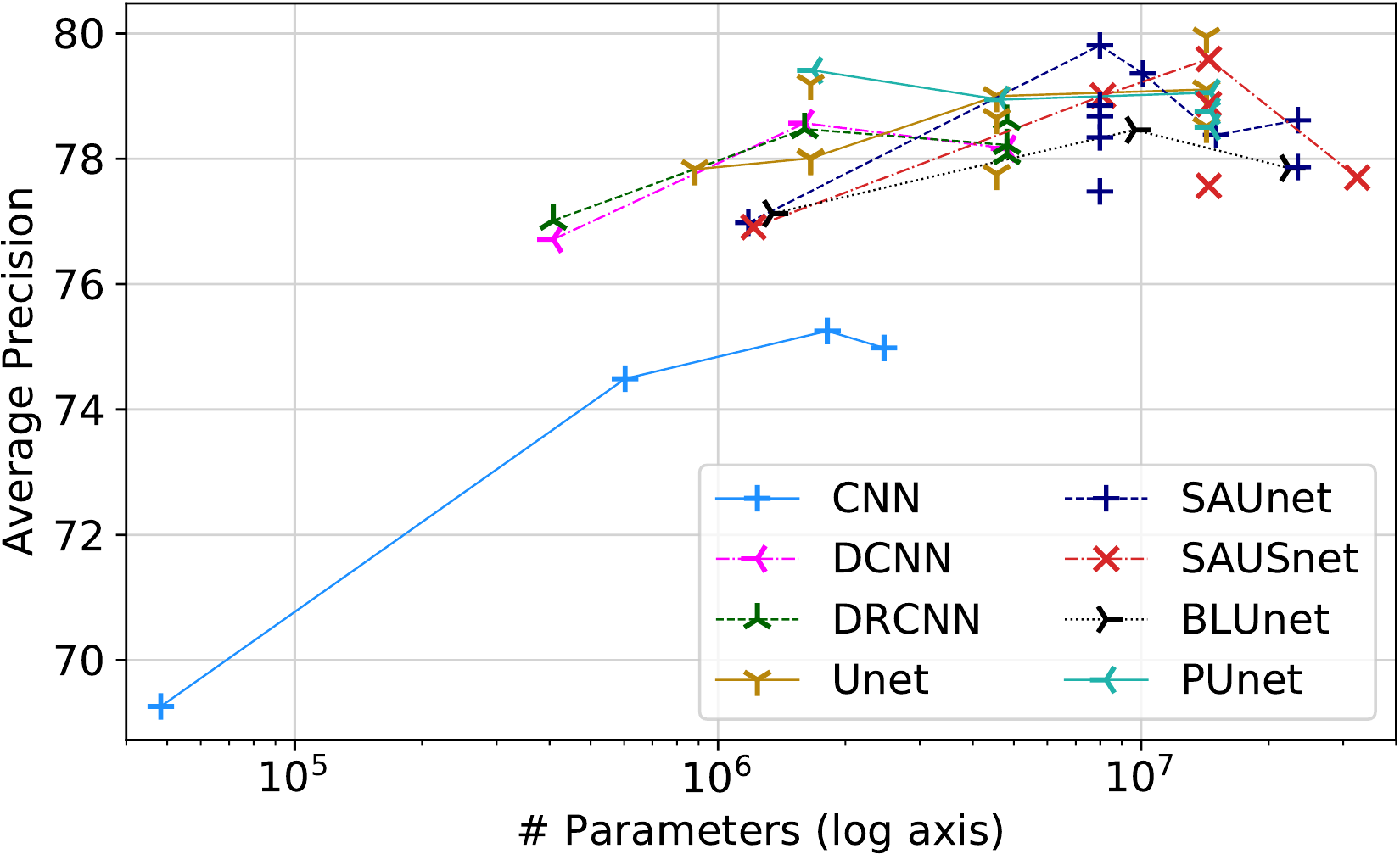}
    \vspace{-0.23cm}
    \caption{Performance of different model types realized in different sizes, specified by their number of paramters. Points connected with line correspond to results reported in \tabref{tab:architecturesmnet}, other points in the same color/style correspond to re-evaluations of the same model. [Best viewed in color.]}
    \label{fig:modelsizes}
\end{figure}

\textbf{Data augmentation.} 
To improve model generalization, we use data augmentation during training (not validation) on the pre-computed HCQT excerpts (in contrast to \cite{ThickstunHFK18_Transcription_ICASSP}, which employs augmentation on the raw waveform). As augmentation strategies, we employ transposition by shifting the HCQT by a random amount of steps (up to $\pm5$ semitones), filling the gaps with zeros, and shifting the labels accordingly. We further employ tuning augmentation by shifting for $\pm1$ CQT bin (1/3 semitone) or by $\pm0.5$ bins (averaging neighboring bins), thereby preserving the labels. Moreover, we add Gaussian noise to all HQCT values with a standard deviation of $10^{-4}$. To simulate different timbral characteristics, we apply the RandomEQ technique presented in \cite{AbesserM21_BassTransUNet_Electronics} with parameters $\alpha\in\{1,\ldots,21\}$ and $\beta\in\{1,\ldots,216\}$ (for details, see \cite{AbesserM21_BassTransUNet_Electronics}). In preliminary experiments, we found data augmentation to have a small but consistently positive effect on test performance.

\textbf{Evaluation measures.} 
To assess the effectiveness of our models, we predict multi-pitch estimates for all test files with a dense sampling (stride of 1 as in \cite{ThickstunHK17_MusicNet_ICLR}), i.\,e., all test frames are predicted and evaluated, at a frame rate of \unit[43.07]{Hz}. As evaluation measures, we compute the Precision (P), Recall (R), and F-Measure (F).\footnote{We use the implementations by \scriptsize\protect \url{https://craffel.github.io/mir_eval/}. For \secref{sec:testsets}, we also report the \ac{acc} score as defined by \cite{PolinerE07_PolyphonicPiano_EURASIP} to enable comparability to related work that only reports \ac{acc}.} To binarize model outputs, we use a global prediction threshold of 0.4 as in \cite{ThickstunHFK18_Transcription_ICASSP} instead of tuning the threshold on the validation set (as done in \cite{WuCS19_PolyphTranscription_ICASSP}). Hypothesizing that a more general, threshold-independent measure is better suitable to estimate model quality, we also compute the \ac{avgprec} score, which corresponds to the Area under the Precision--Recall curve, a concept similar to the Receiver-Operator-Characteristics (ROC) curves. As in  \cite{ThickstunHK17_MusicNet_ICLR,ThickstunHFK18_Transcription_ICASSP}, we use \ac{avgprec} as our primary evaluation measure.\footnote{For \ac{avgprec}, we use the scikit-learn implementation (version\,0.21).} All measures reported in this paper are \emph{class-wise micro-averages} (all active bins contribute equally irrespectively of the pitch) and \emph{track-wise macro-averages} (all test tracks contribute equally irrespectively of their length), given in percent.

 \begin{table*}[t]
\caption{Different test subsets of the MusicNet dataset. $^\ast$\,See footnote 12 for handling Bach's Well-tempered Clavier.} 
\vspace{-.2cm}
\centering
\footnotesize
\setlength{\tabcolsep}{3.2pt}
\renewcommand{\arraystretch}{0.92}
\begin{tabular}{ll|lllll}
\toprule
 & & \multicolumn{5}{c}{Test set tracks (MusicNet ID, movement no. and tempo)} \\
 Composer & Work & \textbf{\testThree} / \testTen & \testTenWrong & \testTenSlow & \testTenFast &\testTenFull \\
\midrule
 Bach & Well-tempered Clavier Book 1 & \textbf{2303, 5-Prel.\,D\,maj} & 2303, 5-Prel.\,D\,maj & 2302, 5-Fugue\,D\,maj  & 2310, 15-Prel.\,G\,maj & 2302--2305\,$^\ast$ \\
 Mozart & Serenade K\,375 & \textbf{1819, 4-Menuetto} & 1819, 4-Menuetto & 1818, 3-Adagio  & 1817, 2-Menuetto & 1817--1819 \\
 Beethoven & String Quartet No.\,13 Op.\,130 & \textbf{2382, 2-Presto} & 2382, 2-Presto & 2383, 3-Andante & 2381, 1-Adagio-Allegro & 2381--2384 \\
 Bach & Cello Suite No.\,4 BWV\,1010 & 2298, 6-Gigue & 2298, 6-Gigue & 2293, 1-Prelude & 2296, 4-Sarabande & 2293--2298 \\
 Bach & Violin Partita No.\,3 BWV\,1006 & 2191, 6-Bourree & 2191, 6-Bourree & 2186, 1-Preludio & 2186, 1-Preludio & 2186, 2191 \\
 Beethoven & Piano Sonata No.\,30 Op.\,109 & 2556, 2-Prestissimo & 2556, 2-Prestissimo & 2557, 3-Gesangvoll & 2555, 1-Vivace-Adagio & 2555--2557 \\
 Beethoven & Wind Sextet Op.\,71 & 2416, 3-Menuetto & 2416, 3-Menuetto & 2415, 2-Adagio  & 2417, 4-Rondo-Allegro & 2415--2417 \\
 Beethoven & Violin Sonata No.\,10 Op.\,96 & 2628, 3-Scherzo & 2629, 4-Poco\,allegretto & 2627, 2-Adagio\,espr. & 2626, 1-Allegro\,mod. & 2626--2629 \\
 Schubert & Piano Sonata D\,958 & 1759, 3-Men.\,Allegro & 1759, 3-Men.\,allegro & 1758, 2-Adagio   & 1757, 1-Allegro & 1757--1760 \\
 Haydn & String Quartet Op.\,645 & 2106, 3-Men.\,Allegretto & 2106, 3-Men.\,allegretto & 2105, 2-Adagio\,cant. & 2104, 1-Allegro\,mod. & 2104--2106 \\
\bottomrule
\end{tabular}
\label{tab:testsets}
\end{table*}

\subsection{Evaluating Model Architectures and Sizes}\label{sec:modelsizes}

As our first experiment, we now evaluate all models proposed in \secref{sec:ownmethods} on the \musicnet dataset (results listed in \tabref{tab:architecturesmnet}). To this end, we use the 10-track test set \testTenWrong as specified in \tabref{tab:testsets} (we discuss the choice of the test set in the following section) and further 27 tracks for validation. For each model type, we realize different sizes by adjusting the parameters N$_0, \ldots,\, $N$_3$ and, if applicable, $\gamma$ and $\lambda$ (see \secref{sec:ownmethods}) with resulting model sizes from roughly 48k (\cnn{XS}) to 32M (\sausnet{XXL}) parameters. We consider the self-attention U-net, \saunet{L}, with roughly 8M parameters, to be most similar to the well-performing model of \cite{WuCS20_TranscriptionSelfattn_TASLP}. In addition, \figref{fig:modelsizes} visualizes the \ac{avgprec} values for the models listed in \tabref{tab:architecturesmnet}, with models of the same type connected by lines (we later discuss the additional disconnected points).

\textbf{Standard CNNs.} 
Let us first regard the results of the shallow \cnnNoArg models (\tabref{tab:architecturesmnet}a). For the smallest model \cnn{XS}, which corresponds to the one tested in \cite{WeissP21_MultiPitchMCTC_WASPAA}, we obtain a reasonable result of \ap=69.3\%. By widening this model (using more channels), we can substantially improve its performance up to \ap=75.3\%. We reach a peak at a model size of roughly 1.8M parameter before the performance slightly drops (compare \figref{fig:modelsizes}). A similar ``peaking'' behavior can be found when we increase the model depth (\dcnnNoArg, \tabref{tab:architecturesmnet}b) and when adding residual connections to this deeper CNN (\drcnnNoArg, \tabref{tab:architecturesmnet}c).\footnote{For the large variants \dcnn{M}, \dcnn{L}, \drcnn{M}, \drcnn{L} to converge, we needed a smaller initial learning rate of 0.0002.} Compared to the shallow \cnnNoArg, the deeper models achieve superior results of \ac{avgprec}$>$78\%, suggesting that a certain depth is necessary for good performance.

\textbf{U-nets.} 
Looking at the U-net models (\tabref{tab:architecturesmnet}d), we observe similar performance for the smaller model sizes (\unet{S} and \unet{M}). Nevertheless, it seems that using a U-net structure allows to substantially increase the model size up to 14M parameters (\unet{XL}) to achieve improved results of \ap=79.1\%. Enhancing the U-net with self-attention modules at the bottom (\saunetNoArg, \tabref{tab:architecturesmnet}e), we find even better results up to \ap=79.8\% for \saunet{L}, the model similar to \cite{WuCS20_TranscriptionSelfattn_TASLP}. Interestingly, performance drops for the larger variants of this model (\saunet{XL} and \saunet{XXL}), again suggesting that an ideal size exists for this model as observed above.

\textbf{Proposed extensions.} 
Regarding our own extensions of the models, let us first discuss the \sausnetNoArg models (\tabref{tab:architecturesmnet}f), where we added self-attention modules also to the U-net's lowest skip connection (in addition to the bottleneck). These models also perform well with the peak result of \ap=79.6\% at a slightly larger size of roughly 14M parameters. Replacing bottleneck self-attention with recurrent modules (\blunetNoArg, \tabref{tab:architecturesmnet}g), we obtain worse results than for \saunetNoArg. Finally, let us turn to the multi-task model \punetNoArg (\tabref{tab:architecturesmnet}h), where we constrain the bottleneck by training on degree-of-polyphony estimation. These models show good performance comparable to the standard U-nets. However, the smallest model with polyphony estimation (\punet{M}) obtains \ap=79.4\%, being superior to the equally-sized \unet{M} with \ap=78.0\%, suggesting that multi-task training can help for smaller models.
Summarizing these results, we can achieve well-performing \ac{mpe} systems with any of our convolutional approaches. To boost performance, a certain model depth seems to be necessary; we obtained further improvements by using U-net architectures and additional enhancement strategies.

\textbf{Reliability of results.} 
At this stage, we want to point out that differences in \ac{avgprec} (and other evaluation measures) are usually small---often less than one percentage point. This raises the question of the results' stability. We therefore re-run some selected model configurations. Since any type of stochastic gradient descent depends on randomization, we expect at least slightly different results. In our case, randomization is involved at four stages of model training: The random initialization of model parameters $\theta$, the random order of training samples and batches in an epoch, the random parameter choices involved in the data augmentation, and the randomness in dropout layers.

We plot the \ac{avgprec} values from the additional re-runs as extra isolated points into \figref{fig:modelsizes} (same marker types at the same horizontal positions stem from identical models). The result is remarkable: For our best model from \tabref{tab:architecturesmnet} (\saunet{L}), each re-run leads to worse results, down to \ap=77.5\%. For the second best model, \sausnet{XL}, re-runs produce worse results as well. In contrast, one re-run of \unet{XL} achieves a top result of \ap=79.9\%. In general, the variation in \ac{avgprec} between runs reaches up to more than two percentage points, which is clearly larger than most performance differences between model architectures and sizes. This not only questions the usefulness of the self-attention components and the other U-net enhancements but also the role of model architecture at all: We cannot even claim U-net-based models to be superior to \dcnnNoArg or \drcnnNoArg architectures from this experiment. However, this does not allow for the opposite conclusion (that these architectures are \emph{not} better)---we just cannot safely conclude anything on model superiority from this experiment.

 
\subsection{Investigating Model Generalization}\label{sec:testsets}
To approach this problem, we reconsider several components of our procedure. Hypothesizing that a larger training and validation set may stabilize model performance across runs, we increase their size by a factor of 2.5 using denser HCQT sub-sampling (smaller stride), raising the number of training examples from roughly 95k to 236k and the number of validation examples from roughly 9k to 23k.\footnote{To leave the training procedure (scheduling and early stopping) unchanged, we declare an epoch as finished after 3800 batches (95k examples), thus covering the same amount of batches as in the previous experiment.}

\textbf{Larger training set.} 
\tabref{tab:moresamplesandtestsets}a shows the results of this adapted experiment for selected models. For the three \saunet{L} runs, we observe performance to stabilize (only 0.4 percentage points variation) and to improve (\ac{avgprec}$>$80\% for two runs). This improvement is confirmed for \sausnet{XL} and \punet{XL}, which surpass their performance with less training data in \tabref{tab:architecturesmnet}. However, the runs of \unet{XL} in \tabref{tab:moresamplesandtestsets} show a variation of more than three percentage points in \ac{avgprec}, indicating that an increase of training data is not sufficient for safely excluding negative randomization effects.

 \begin{table}[t]
\caption{Results with larger training set (denser sampling) and (partially) re-runs, evaluated with different test sets.} 
\vspace{-.2cm}
\centering
\footnotesize
\setlength{\tabcolsep}{3.4pt}
\renewcommand{\arraystretch}{0.9}
\begin{tabular}{lrl|lllll}
\toprule
 Model & Param. & Run & \multicolumn{5}{c}{Eval. Measures (\%)} \\
 Type:Size & $/10^3$ &  & P & R & F & \ac{avgprec} & \ac{acc} \\ 
\midrule
\multicolumn{8}{c}{\textbf{(a) Test set} \testTenWrong ~(\secref{sec:modelsizes})} \\
\midrule
\unet{XL} & 14,252 & 1 & 68.3 & 80.0 & 73.3 & 77.2 & 58.3 \\
          &        & 2 & 71.9 & 77.3 & 74.2 & 78.3 & 59.3 \\
          &        & 3 & 76.0 & 74.9 & 75.2 & 80.3 & 60.6 \\[3.4pt]
\saunet{L} & 7,983 & 1 & 75.8 & 73.4 & 74.3 & 80.0 & 59.7 \\
           &       & 2 & 74.6 & 76.6 & 75.4 & 80.1 & 60.8 \\
           &       & 3 & 74.5 & 74.7 & 74.4 & 79.7 & 59.5 \\[3.4pt]
\sausnet{XL} & 14,436 & -- & 75.3 & 76.3 & 75.6 & 80.2 & 61.1 \\[3.4pt]
\punet{XL} & 14,598 & -- & 75.1 & 75.6 & 75.1 & 80.6 & 60.4 \\[3.4pt]
\midrule
\multicolumn{8}{c}{\textbf{(b) Test set} \testTen \cite{ThickstunHFK18_Transcription_ICASSP,WuCS19_PolyphTranscription_ICASSP}} \\
\midrule
\unet{XL} & 14,252 & 1 & 74.8 & 73.1 & 73.7 & 78.8 & 58.7 \\
           &       & 2 & 73.6 & 75.3 & 74.2 & 79.2 & 59.4 \\
           &       & 3 & 73.2 & 78.1 & 75.5 & 80.4 & 61.0 \\[3.4pt]
\saunet{L} & 7,983 & 1 & 72.2 & 76.8 & 74.2 & 79.0 & 59.7 \\
           &       & 2 & 72.9 & 72.7 & 72.6 & 77.4 & 57.4 \\
           &       & 3 & 73.1 & 76.0 & 74.3 & 79.5 & 59.7 \\[3.4pt]
\sausnet{XL} & 14,436 &  1 & 69.8 & 79.4 & 74.1 & 78.7 & 59.1 \\
             &        &  2 & 72.6 & 78.0 & 75.0 & 79.6 & 60.4 \\
             &        &  3 & 71.2 & 78.0 & 74.2 & 78.8 & 59.3 \\[3.4pt]
\punet{XL} & 14,598 & -- & 75.6 & 74.7 & 74.9 & 80.6 & 60.2 \\[3.4pt]
\midrule
\multicolumn{8}{c}{\textbf{(c) Test set} \testThree~(90s) \cite{ThickstunHK17_MusicNet_ICLR,ThickstunHFK18_Transcription_ICASSP}} \\
\midrule
\unet{XL} & 14,252 & -- & 66.3 & 76.4 & 70.8 & 74.6 & 55.4 \\[3.4pt]
\saunet{L} & 7,983 & 1 & 74.5 & 64.1 & 68.8 & 75.0 & 53.3 \\
           &       & 2 & 74.1 & 74.0 & 74.0 & 78.4 & 59.2 \\
           &       & 3 & 74.8 & 70.5 & 72.5 & 78.0 & 57.6 \\[3.4pt]
\sausnet{XL} & 14,436 & -- & 74.8 & 71.7 & 73.2 & 77.5 & 58.4 \\[3.4pt]
\punet{XL} & 14,598 & -- & 74.5 & 69.4 & 71.8 & 77.3 & 56.5 \\[3.4pt]
\midrule
\multicolumn{8}{c}{\textbf{(d) Test set} \testTenSlow ~(slow movements)} \\
\midrule
\saunet{L} & 7,983 & -- & 75.2 & 82.1 & 78.5 & 83.0 & 64.8 \\[3.4pt]
\midrule
\multicolumn{8}{c}{\textbf{(e) Test set} \testTenFast ~(fast movements)} \\
\midrule
\saunet{L} & 7,983 & -- & 75.0 & 72.2 & 73.4 & 78.0 & 58.1 \\[3.4pt]
\midrule
\multicolumn{8}{c}{\textbf{(f) Test set} \testTenFull ~(all movements)} \\
\midrule
\cnn{M} & 1,813 & -- & 74.3 & 72.7 & 73.2 & 78.3 & 58.0 \\[3.4pt]
\drcnn{L} & 4,815 & 1 & 73.2 & 79.4 & 76.0 & 80.6 & 61.5 \\
           &       & 2 & 72.3 & 81.1 & 76.2 & 80.8 & 61.8 \\
           &       & 3 & 74.2 & 79.3 & 76.5 & 81.2 & 62.1 \\[3.4pt]
\unet{M} & 1,655 & -- & 74.7 & 75.8 & 75.1 & 80.1 & 60.4 \\[3.4pt]
\unet{XL} & 14,252 & -- & 74.5 & 78.3 & 76.2 & 80.9 & 61.8 \\[3.4pt]
\saunet{L} & 7,983 & 1 & 76.2 & 74.6 & 75.2 & 80.4 & 60.5 \\
           &       & 2 & 75.8 & 75.7 & 75.6 & 80.5 & 60.9 \\
           &       & 3 & 73.6 & 78.7 & 75.9 & 80.7 & 61.3 \\[3.4pt]
\sausnet{XL} & 14,436 & -- & 72.0 & 79.8 & 75.5 & 80.1 & 60.9 \\[3.4pt]
\blunet{L} & 9,649 & -- & 76.7 & 72.7 & 74.5 & 79.8 & 59.6 \\[3.4pt]
\punet{XL} & 14,598 & 1 & 76.2 & 75.5 & 75.7 & 80.8 & 61.1 \\
           &        & 2 & 80.0 & 70.6 & 74.8 & 81.5 & 60.1 \\
           &        & 3 & 76.6 & 76.9 & 76.5 & 81.4 & 62.2 \\
\bottomrule
\end{tabular}
\label{tab:moresamplesandtestsets}
\end{table}

\textbf{MusicNet test sets in related work.} 
As another critical aspect, we want to reflect on the test set held out from \musicnet for evaluation. Studying related work on \musicnet, we observe different test sets being used (compare \tabref{tab:testsets}). In the original paper \cite{ThickstunHK17_MusicNet_ICLR} and a follow-up work by its authors \cite{ThickstunHFK18_Transcription_ICASSP}, a test set of three tracks (\testThree) is used, restricted to the first 90 seconds (which are further subsampled), taken up by some other authors \cite{PedersoliTY20_TranscriptionPreStackUNet_ICASSP,CheukLBH20_SpecReconstructionAMT_ICPR}. In a pre-print\footnote{This version deviates from the one officially published at ICASSP and is hosted on a pre-print server under {\protect \url{https://arxiv.org/abs/1711.04845}}, \textbf{[v1]}.} of \cite{ThickstunHFK18_Transcription_ICASSP}, Thickstun \etal further evaluate on a ten-track test set (\testTen), extending \testThree by seven further tracks (now full tracks), with remarkable differences (\ap=77.3\% for \testThree and \ap=79.9\% for \testTen with their best model). A number of subsequent studies evaluates on \testTen \cite{WuCS19_PolyphTranscription_ICASSP,WuCS20_TranscriptionSelfattn_TASLP,SteinerSBJ20_MultipitchEchoStateNetworks_EUSIPCO} or an entirely different, customized training--test split \cite{GardnerSMHE21_MultiTaskTranscription_arXiv}.

\textbf{A small ``mistake.''} 
In our own previous experiments (\secref{sec:modelsizes}), we targeted a ten-track test set inspired by the literature. However, we later realized that one of our test tracks deviates from \testTen due to a mistyped ID (2629 instead of 2628)---this is why we named it differently as \testTenWrong (see \tabref{tab:testsets}). To enable comparability with published results, we therefore re-train and test some of our models on the corrected test set \testTen (see \tabref{tab:moresamplesandtestsets}b). Interestingly, we obtain worse results than in \tabref{tab:moresamplesandtestsets}a especially for \saunet{L} and \sausnet{XL}, with the best result on \testTen, \ap=79.6\%, being worse than the worst result of \testTenWrong, \ap=79.7\%. Only one run of \unet{XL} achieves better results with \ap=80.4\%, and the multi-task \punet{XL} performs similar. Testing on the original test set \testThree from \cite{ThickstunHK17_MusicNet_ICLR} (first 90s) as shown in \tabref{tab:moresamplesandtestsets}c, we observe substantially worse results than for any ten-track test set, in accordance with \cite{ThickstunHFK18_Transcription_ICASSP}.

Given these results, we now could follow the common practice and claim ``state-of-the-art'' results for several of our models, without even limiting to a particular test set or evaluation measure: On the test set \testTen (\tabref{tab:moresamplesandtestsets}b), \punet{XL} (\ap=80.6\%) and \unet{XL} in run 3 (\ap=80.4\%) outperform \cite{ThickstunHFK18_Transcription_ICASSP} (\ap=79.9\%) in \ap, and almost all our models outperform \cite{WuCS19_PolyphTranscription_ICASSP} (F=73.3\%) and \cite{SteinerSBJ20_MultipitchEchoStateNetworks_EUSIPCO} (F$<$72\%) in F-measure. On the test set \testThree (\tabref{tab:moresamplesandtestsets}c), our \saunet{L} in run 2 (\ap=78.4\%) outperforms the models from \cite{ThickstunHFK18_Transcription_ICASSP} (\ap=77.3\%), from \cite{PedersoliTY20_TranscriptionPreStackUNet_ICASSP} (\ap=76.8\%), and from \cite{CheukLBH20_SpecReconstructionAMT_ICPR} (\ac{avgprec}$<$71\%). However, \emph{we strongly believe that such comparisons are not meaningful} given the small margin of improvement compared to the large variation of results across runs and chosen test sets.

\begin{figure}[t]
    \centering
    \includegraphics[width=.85\columnwidth]{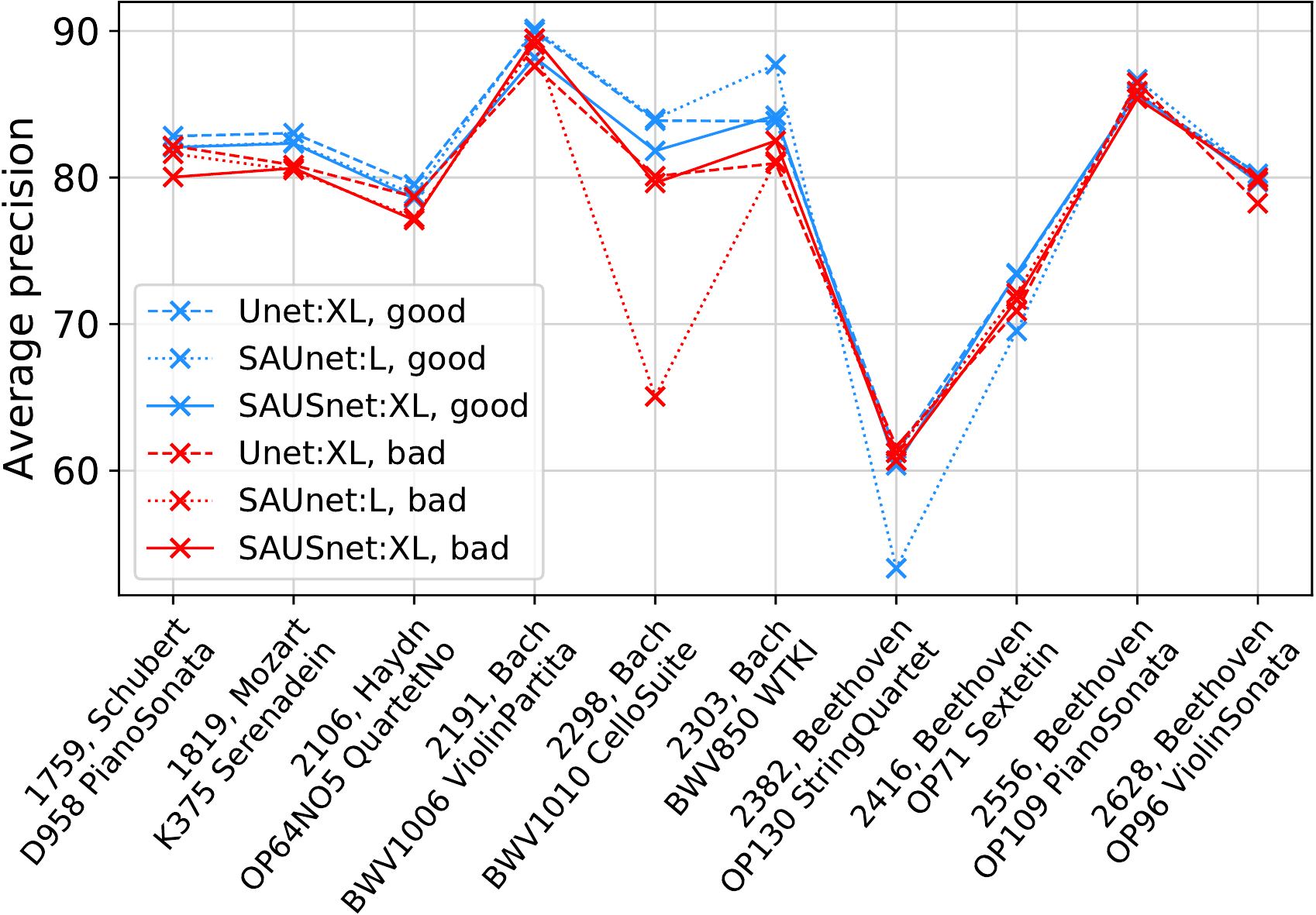}
     \vspace{-0.35cm}
    \caption{Results on the individual test files for \testTen with three good and three bad training runs.}
    \label{fig:testpiece_sweep}
\end{figure}

\textbf{Results in detail.} 
Since there seems to be a substantial effect of the test set, we investigate the behavior for individual test tracks. For each of the three models from \tabref{tab:moresamplesandtestsets}b, we take the ``good'' and the ``bad'' training run and plot the track-wise \ac{avgprec} values in \figref{fig:testpiece_sweep}. 
For some tracks (IDs 1759, 1819, 2106, or 2303), good models are consistently but slightly better than bad models. Others achieve consistently high (IDs 2191 or 2556) or low (ID\,2416) \ac{avgprec} throughout all models and runs. Two test tracks are behaving quite differently: For ID 2298, a movement from Bach's Cello Suite BWV\,1010, the performance substantially drops for \saunet{L}. For ID 2382, a very fast movement from Beethoven's string quartet op.\,30, all models show worse results. Looking in detail at this track's annotations (see \figref{fig:mnetexample} or consult \protect \url{https://musicnet-inspector.github.io/}), we realize synchronization issues for this track, which---due to the fast tempo---result in note events being shifted to their neighbors in time. Such kind of annotation problems were highlighted for \musicnet by several authors. Already in the original paper \cite{ThickstunHK17_MusicNet_ICLR}, the authors report a quite high onset/offset error rate of 4\%. Beyond that, Gardner \etal \cite{GardnerSMHE21_MultiTaskTranscription_arXiv} found timing errors beyond \unit[50]{ms} 
and spotted major labeling errors due to non-corresponding score--audio segments.\footnote{See ID\,1817 at \protect\url{https://musicnet-inspector.github.io/} for an example.}

\textbf{Impact of test set choice.} 
Beyond such labeling problems, there is another important aspect of training--test splits that we already discussed in \secref{sec:introduction}. Our small typo (ID 2629 instead of 2628) did not result in an entirely different test track being used but only in a different movement (4-Poco\,allegretto instead of 3-Scherzo) of the same cycle (Beethoven's violin Sonata No.\,10 Op.\,96) in the same version (same performers and recording scenario) being used. This similiarity is no exception in \musicnet, which constitutes a mixed-version datasets as highlighted in \figref{fig:teaser}d: Each cycle--version pair usually contributes several works (movements) to the dataset (organized in separate audio tracks, each with their own \musicnet ID). This has two consequences: First, models can overfit to the specific recording conditions of the test set, 
which can result in unrealistically high model performance. Second, musical aspects correlated to the movement type or tempo may influence the results, which might be one reason for the performance difference between \testTen and \testTenWrong.

\textbf{Switching test movements.} 
To highlight the impact of the latter issue, we now create another ten-track test set, where we replace each of the original test tracks from \testTen with \emph{another movement from the same cycle in the same version}. Since \testTen contains mostly fast movements, we now take mostly slow movements, resulting in the new test set \testTenSlow (compare \tabref{tab:testsets}). Retraining the model \saunet{L} with this training--test split (\tabref{tab:moresamplesandtestsets}d), we obtain clearly improved performance of \ap=83.0\% (note that the performance difference between this result and any previous one is larger than the training run differences found so far). We repeat this process by taking yet another movement (now mostly fast ones as in \testTen) resulting in the test set \testTenFast. Here, the model \saunet{L} achieves \ap=78.0\% (\tabref{tab:moresamplesandtestsets}e), which is quite similar to the result for \testTen, suggesting that tempo-related characteristics can indeed influence model performance. 
This might be due to the lower average note length in fast movements, which results in synchronization errors having more influence on annotation quality.

\textbf{Proposed training--test split.} 
To address both the tempo and the generalization issues, we finally propose a stricter split where we again take the same ten work cycles as in \testTen but hold out \emph{all movements of each cycle} for testing, resulting in 36 tracks (\testTenFull, last column in \tabref{tab:testsets}). This is beneficial for three reasons: First, the test set becomes larger, thus suppressing the impact of individual difficult (or misannotated) tracks as highlighted in \figref{fig:testpiece_sweep}. Second, the influence of tempo and movement characteristics is balanced out since, for each cycle, we consider a variety of movements. Third, we avoid overfitting to the specific acoustic characteristics of a recording since, for each recorded cycle, all movements are either in the training or test set.\footnote{We remove all movements from Bach's Well-tempered Clavier (the 39 tracks labeled ``WTKI'') but use only four of them for testing (IDs 2302--2305) in order to prevent these piano tracks from dominating the test set.} We propose this training--test split as an improved evalution strategy for \musicnet, which largely avoids the problems mentioned above.

The results for this split are shown in \tabref{tab:moresamplesandtestsets}f. We observe slightly better results than for \testTen and somewhat worse results than for \testTenSlow, indicating that the test set \testTen proposed in the appendix of \cite{ThickstunHFK18_Transcription_ICASSP} seems to be a representative choice. However, in contrast to choosing the smaller variant of ten tracks, we now observe clearly stabilized results. For the three models with re-runs (\drcnn{L}, \saunet{L}, and \punet{XL}), the variation between best and worst run is at most 0.7 percentage points, which is clearly lower than in \figref{fig:modelsizes}. Comparing model architectures on \testTenFull confirms most of our initial findings: First, a certain depth of the network seems necessary---the shallow \cnn{M} performs worse than deeper models. Second, U-nets with self-attention do not prove to be better than standard U-nets of comparable size. Third, a larger U-net can outperform a smaller one. Fourth, a simple residual network (\drcnn{L}) can perform equivalently to a U-net. Fifth, recurrent components are less effective than self-attention components, rather deteriorating results compared to a standard U-net. Finally, we find evidence that the multi-task strategy (\punet{XL}) is a promising technique.

\subsection{Cross-Version Study: Schubert Winterreise Dataset}\label{sec:crossversion}

\begin{figure}[t]
    \centering
    \includegraphics[width=.8\columnwidth]{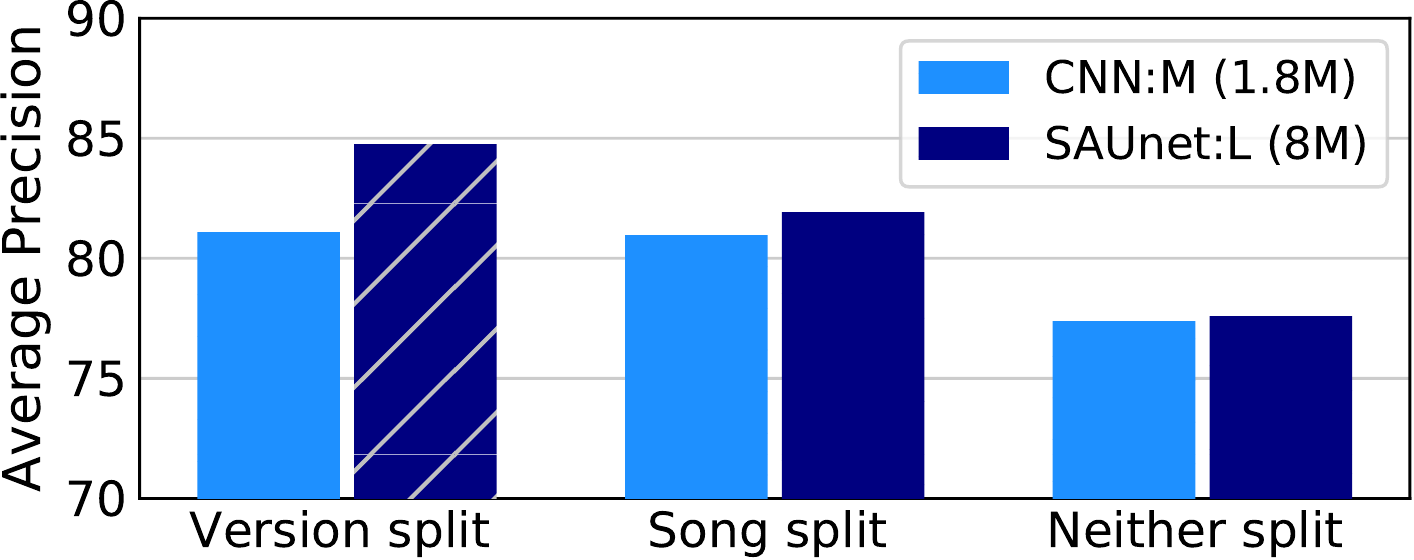}
     \vspace{-0.3cm}
    \caption{Cross-version study on \schubert inspired by \cite{WeissSM20_LocalKey_TASLP} for two selected models, evaluated in a version split, a song split, and an exclusive neither split.}
    \label{fig:crossversion_schubert}
\end{figure}

Our previous experiments showed that work-specific aspects such as the tempo can have major influences on model performance. Yet, these experiments provided little insights about overfitting to certain \emph{versions} (performance and recording conditions). To test this in a more controlled fashion, we turn to the \schubert \cite{WeissZAGKVM21_WinterreiseDataset_ACM-JOCCH}, a systematic multi-version dataset, which is \emph{comprehensive} in the sense that each work is present in all versions and each version covers all works (\figref{fig:teaser}c). The dataset has been used for a cross-version study on local key estimation \cite{WeissSM20_LocalKey_TASLP}, where models rather overfit to works (\emph{songs}) than to versions. 
%
We now realize a similar cross-version study for \ac{mpe}. 
For the \emph{version split}, we use all songs in two versions for testing, two further versions for validation, and the remaining five versions for training. For the \emph{song split}, we use the songs 17--24 of \emph{Winterreise} in all versions for testing, songs 14--16 for validation, and songs 1--13 for training. The neither split is the intersection of the two splits.

\figref{fig:crossversion_schubert} shows the results of this study for two selected models, the shallow \cnn{M} and the large \saunet{L}. 
Similar to \cite{WeissSM20_LocalKey_TASLP}, the larger model (\saunet{L}) benefits from knowing the specific songs. For \cnn{M}, \emph{version} and \emph{song split} behave similar. For the most realistic \emph{neither split}, where neither the songs nor the acoustic conditions are seen by the model during training, both models suffer from a substantial drop (please note the y-scale: differences here are larger than in the previous sections). We conclude on the existence of a \emph{version effect} (similar to the ``album effect'' in \cite{Flexer07_GenreClassification_ISMIR}) for both models and a \emph{song effect} (similar to the ``cover song effect'' in \cite{WeissSM20_LocalKey_TASLP}) in particular for the high-capacity model \saunet{L}.

In general, we note that performance in the realistic \emph{neither split} is worse than for \musicnet, even though generalization is less challenging since we only deal with one composer, one work cycle, and one instrumentation. We suppose that this is due to the limited amount and variety of training data in \schubert compared to \musicnet, which prohibits high-capacity models such as \saunet{L} from fully exploiting their potential.

\subsection{Evaluation With a Mixed Dataset}\label{sec:crossdataset}

\begin{figure*}[t]
    \centering
    \includegraphics[width=\textwidth]{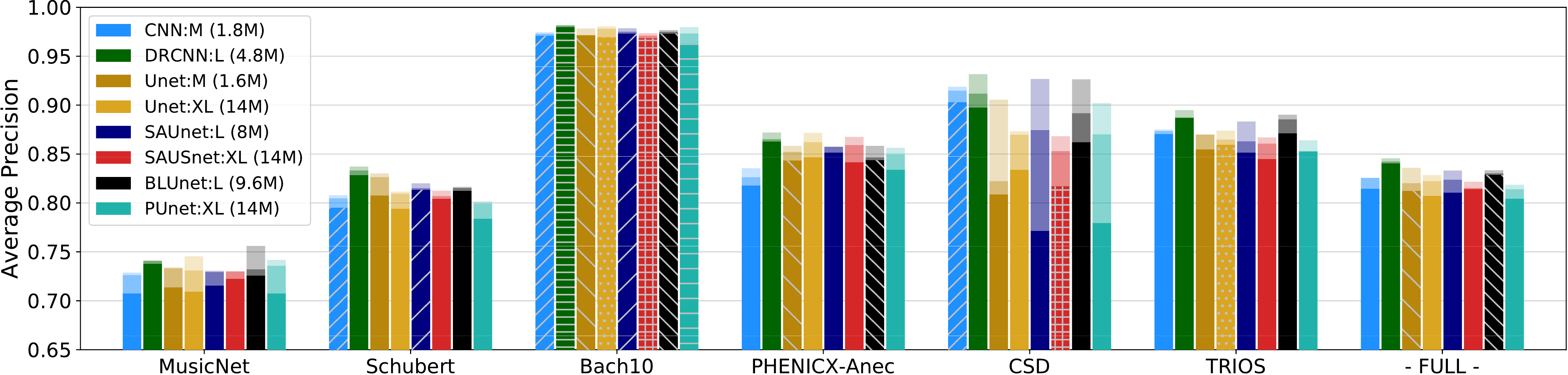}
    \vspace{-0.7cm}
    \caption{Generalization study on a big mixed dataset. Results of different training runs are stacked on top of each other with transparent colors.}
    \label{fig:cross_dataset}
\end{figure*}

To address the challenge of training dataset size and variety for general \ac{mpe}, we finally compile a big mixed dataset using all the datasets listed in \tabref{tab:datasets} as sources. 
For \musicnet, we use the test set \testTenWrong as in \secref{sec:modelsizes}. For \schubert, we employ the strict \emph{neither} split as in \secref{sec:crossversion}. We use the full \trios dataset for testing since it comprises chamber music recordings very similar to \musicnet (but with very high annotation quality). All other datasets are splitted.\footnote{For the detailed splits, please consult our code repository.}
Since the \choralsd is our only choir dataset, we enrich it by creating for each work also a three-part mix (once leaving out each of the four parts, respectively).
The combined test set comprises 42 tracks and, thus, is comparable in size to \testTenFull. Since the training set is much larger, we increase the sampling stride so that the number of training and validation examples is comparable to \secref{sec:testsets}. 

\textbf{Results on the full test set.} 
Using this combined training--test scenario, we evaluate several models, each in three runs. \figref{fig:cross_dataset} shows the results, with the three runs' \ac{avgprec} values stacked on top of each other with transparent colors. Looking at the full mixed test set (rightmost bar group), we observe high \ac{avgprec} values of 80--85\% for all models. Interestingly, the residual model \drcnn{L} performs best, with its worst run obtaining better \ac{avgprec} than any run of any other model. Again, self-attention components do not improve performance of U-nets. Contrary to the findings for \musicnet, the recurrent components in \blunet{L} are beneficial, the smaller \unet{M} is slightly better than the larger \unet{XL}, and the multi-task \punet{XL} performs worse---yet on an overall high level.

\textbf{Results on test subsets.} 
Much more interesting, however, are the results for the individual test subsets. For \schubert, all models show better performance using this large training set than for the corresponding \emph{neither split} in \figref{fig:crossversion_schubert}. This confirms our previous assumption that a larger training set is beneficial here. On the contrary, the results for \musicnet drop by more than five percentage points compared to the corresponding results in \tabref{tab:moresamplesandtestsets}a. Remarkably, a more diverse training set from different sources does not help for \musicnet (in contrast to \schubert). We suppose that this is due to specific annotations characteristics of \musicnet, probably stemming from the synchronized annotations, which the models can effectively adapt to---as long as they only need to model training examples from \musicnet and not at the same time focus on singing voices or orchestral scenarios. For \bach10, a clean and well-annotated dataset, all models yield very good results. \phenicx is a more difficult one since orchestra recordings are underrepresented in our training set. The same holds for \choralsd since we have little choir examples for training. But here, the model performance varies dramatically across models and runs, with the simpler models \cnn{M} and \drcnn{L} performing best. Finally, \trios is the most interesting test set since it is completely excluded from training, well-annotated, and mainly captures our central genre, classical chamber music.\footnote{There is only one jazz recording (\emph{Take Five}) that also includes drums.} Despite being small (\unit[3]{min} overall length), results show high stability across model runs. Interestingly, the performance differences between models are similar to the full test set (though on a higher level). We conclude that \trios is a representative and useful dataset for \ac{mpe} evaluation, and that residual models such as \drcnn{L} seem to be most stable in cross-dataset situations.

\textbf{Comparison with related work.} 
We finally want to stress again that all these models perform well, most of them superior to previously reported results: All models outperform our previous results \cite{WeissP21_MultiPitchMCTC_WASPAA} with a smaller network similar to \cnn{XS}, especially for \schubert. For \bach10, our best  \unet{XL} reaches \ac{acc}=87.2\% as compared to a traditional method \cite{DuanPZ10_MultiF0_TASLP} with \ac{acc}$<$70\% and a deep-learning approach \cite{Elowsson18_DeepMultiPitch_arXiv} with up to \ac{acc}=85.6\%. On \phenicx, \drcnn{L} and \unet{XL} reach up to \ap=87.2\%, as compared to our own MCTC-based results of \ap=82.8\% \cite{WeissP21_MultiPitchMCTC_WASPAA}. For \choralsd, we achieve F=86.2\% with the best \drcnn{L}, as compared to 
a CNN-based approach \cite{CuestaGML18_ChoirIntonation_ICMPC} with F$<$85\% (for \ac{mfe} with \unit[100]{cent} resolution). Finally, on \trios, our best \drcnn{L} achieves up to F=80.4\% as compared to another deep-learning method \cite{Elowsson18_DeepMultiPitch_arXiv} with F=71.8\%.

\section{Discussion}\label{sec:discussion}
Looking at our findings for \musicnet, \schubert, and the mixed dataset, it remains difficult to draw a general conclusion on model architectures. Our most controlled experiment on \musicnet (\testTenFull, \tabref{tab:moresamplesandtestsets}f) indicated that U-net-based models and, in particular, the multi-task strategy \punetNoArg are useful or at least competitive. In contradiction to this, the generalization on our big mixed dataset questioned the superiority of large U-nets in favor of the smaller \drcnn{L}. 
Nevertheless, all of these models are useful and achieve competitive performance on any of the six datasets, comprising a variety of musical and acoustic scenarios. This means that deep-learning approaches based on convolutional architectures and HCQT input constitute an effective and robust approach to general-purpose \ac{mpe}.

As our most important finding, our experiments showed how careful we have to be when drawing conclusions from individual models and training runs, especially when using small and biased datasets. By just ``cherry-picking'' some of our best model runs, we could argue to outperform all previously published results on \emph{any} test set and evaluation measure. Even worse, by only showing partial results, we could construct a variety of ``scientific messages,'' regarding, e.\,g., the relationship between model size and performance, or the benefit of complex network architectures. Because of this, we strongly encourage a higher sensitivity when claiming ``state-of-the-art performance'' on \ac{mpe} and other \ac{mir} tasks based on small improvements---we just do not have datasets of sufficient size and quality given the complexity of the tasks and the variety of music scenarios \cite{BenetosDDE19_TranscriptionOverview_SPM}.

Regarding this dataset issue, we want to stress the importance of a high data quality, in particular regarding the accuracy of annotations. We strongly agree with \cite{GardnerSMHE21_MultiTaskTranscription_arXiv} stating for \musicnet: ``we are not aware of scholarly work which has formally investigated this important issue to date. We encourage further investigation into labeling issues [...].'' Moreover, we pointed out the challenge of selecting a suitable test subset from \musicnet due to movement characteristics and version duplicates between training and test sets. We therefore propose to use the \testTenFull test set, which is larger, more diverse in movements, and less prone to version overfitting, as a best-practice split for \musicnet. 
Furthermore, we encourage researchers to conduct mixed- and cross-dataset experiments similar to \secref{sec:crossdataset} for assessing generalization capabilities in a realistic way. Concretely, we recommend testing models trained on \musicnet also on \trios, which is small but well-annotated and essentially comprises the same musical styles. Further high-quality datasets, especially for complex music scenarios such as orchestral works or styles beyond Western classical music are highly necessary to further improve \ac{mpe} systems.

Finally, we want to emphasize that differences among our models and to previous work are generally small, which is also due to the fact that \ac{mpe} is already a well-researched task. Beyond the frame-wise problem of \ac{mpe}, most remarkable improvements today are happening in other, more challenging transcription tasks, which account for note and instrument information \cite{HawthorneSRSHDE19_MAESTRO_ICLR,HawthorneSSME21_PianoTranscrTransformers_ISMIR,GardnerSMHE21_MultiTaskTranscription_arXiv,WuCS20_TranscriptionSelfattn_TASLP}. In accordance with \cite{WuCS20_TranscriptionSelfattn_TASLP}, we suppose that our U-net-based models might be more effective for such tasks (as compared to, e.\,g., \drcnnNoArg) than for the simpler \ac{mpe} task. Moreover, our paper concentrates on optimal \ac{mpe} \emph{performance}, ignoring the benefits of smaller model size or real-time capabilities. For practical application, small improvements by the large models of this paper might be less important than having a lightweight model with lower but still acceptable performance as in \cite{WeissP21_MultiPitchMCTC_WASPAA}.

\section{Conclusion}\label{sec:conclusion}
We approached the task of general, instrument-agnostic frame-wise music transcription (\ac{mpe}) and tested a variety of deep-learning architectures, mostly based on the U-net paradigm. We proposed several extensions to such U-nets, with a multi-task strategy for simultaneous degree-of-polyphony estimation turning out most promising. As our main contribution, we showed that the superiority of models has to be judged very carefully since improvements are usually small and variation across different training runs and test sets commonly exceeds the marginal improvement achieved by a better model. To this end, we propose a novel training--test split for \musicnet and encourage researchers to conduct cross-dataset studies.


\section*{Acknowledgments}
C.\,W. is funded by the German Research Foundation (DFG WE\,6611/1-1). 
The authors thank Jakob Abe{\ss}er, Ond\v{r}ej C\'ifka, and Kilian Schulze-Forster for fruitful discussions.



\bibliographystyle{IEEEtran}

\end{document}